\documentclass[aps,prd,twocolumn,groupedaddress,showpacs]{revtex4}
\usepackage{graphics}

\begin{document} 


\title{Cosmological Gravitomagnetism and Mach's Principle}


\author{Christoph~Schmid}
\email[]{chschmid@itp.phys.ethz.ch}

\affiliation{Institut f\"ur Theoretische Physik, ETH-H\"onggerberg, 
CH-8093 Z\"urich}

\date{\today}

\begin{abstract}
The spin axes of gyroscopes experimentally define local non-rotating frames,
i.e. the time-evolution of axes of inertial frames.
But what physical cause governs the time-evolution of gyroscope axes?
We consider linear perturbations of Friedmann-Robertson-Walker (FRW) 
cosmologies with $k=0,$ i.e. spatially flat.
We ask: Will cosmological vector perturbations 
(i.e. vorticity or rotational perturbations)
       {\it exactly drag} 
the spin axes of gyroscopes relative to 
the directions of geodesics to quasars in the asymptotic 
unperturbed FRW space?
Using Cartan's formalism with local orthonormal bases
we cast the laws of linear cosmological gravitomagnetism 
into a form showing the
close correspondence with the laws of ordinary magnetism. 
Our results,
valid for any equation of state and 
any form of the energy-momentum tensor
for cosmological matter, are:
1) \nolinebreak The dragging of a gyroscope axis 
by rotational perturbations of matter
beyond the $H$-dot radius 
from the gyroscope is 
       {\it exponentially suppressed},
where $H$ is the Hubble rate, 
and dot is the derivative with respect to cosmic time.
2) \nolinebreak If the perturbation of matter is a homogeneous 
rotation inside some radius around a gyroscope, then 
exact dragging
of the gyroscope axis by the rotational perturbation 
is reached exponentially fast 
as the rotation radius gets larger than the H-dot radius.
3) \nolinebreak For the most general linear cosmological perturbations
the time-evolution of all
gyroscope spin axes 
and the axis directions of all 
local inertial frames 
exactly follow 
a weighted average of the rotational motion of cosmological matter,
i.e. there is 
exact frame-dragging everywhere.
The weight function is the density of
measured angular momentum of matter times $(1/r)$ 
times the Yukawa force $(-d/dr)[(1/r) \exp (- \mu r)],$ 
where $r$ is the geodesic distance from the source to the gyroscope.
The exponential cutoff is given by $\mu^2 = -4 (dH/dt).$ 
Except for the Yukawa cutoff the weight function is the same as in the 
integrated form of Amp\`{e}re's law.---
Our results demonstrate 
(in first-order perturbations of any type 
for FRW cosmologies with $k = 0$)
the validity of  Mach's hypothesis
that axes of local non-rotating
frames precisely follow an average of the motion of cosmic matter.
\end{abstract}

%
%

\maketitle


\section{Introduction}

\subsection{The observational fact}


In tests of general relativity in the solar system
two types of things are compared:
1) On the one hand
measurements of the precession of 
perihelia or of gyroscopes' spin axes
in Gravity Probe B
    \cite{GP.B}
{\it relative to distant stars and quasars.}
2) On the other hand the predictions from the 
solutions of Einstein's equations 
for the solar system in asymptotic Minkowski space, 
which {\it do not contain distant stars or quasars}.
In this comparison the assumption is made (often not spelled out)
that in the solution of Einstein's equations for the solar system
the asymptotic Minkowski space 
is nonrotating relative to distant stars and quasars.
This implicit assumption is tested to high accuracy
by the comparison of the observed perihelion precession 
with the prediction of general relativity. 
This implicit assumption is a basic observational fact,
which has been called ``Mach~0'' e.g. in
    \cite{Bondi}.---
Distant stars and quasars have proper motions 
relative to the uniform Hubble flow, 
but the angular parts are unmeasurable today.---  
We conclude that the measurements of perihelion shifts  
and the measurement undertaken by Gravity Probe B are 
    {\it tests of two things combined}, 
on the one hand tests of Einstein's equations,
on the other hand tests of the principle ``Mach~0''.


{\it No explanation} for the observational fact ``Mach~0'' is given in
classical mechanics, special relativity, and 
general relativity  for isolated systems in asymptotic Minkowski space, 
except by invoking an accident of initial conditions. 
In these theories one could have different initial conditions,
where distant stars and quasars could be orbiting 
around us relative to our gyroscopes.---
Within these three theories (and in the generic case of general relativity) 
the local non-rotating frames (axis directions of inertial frames)
can be experimentally determined by axes of gyroscopes.
Conversely the time evolution of gyroscope axes is
dictated by the laws of inertia, i.e. the gyroscope
axes cannot rotate with respect to local inertial axes.
This is self-consistent: 
One uses two gyroscopes (with spin axes not aligned) 
to experimentally construct, i.e. 
     {\it operationally define,} 
the local non-rotating frame,
and one uses all other local gyroscopes and other experiments 
(e.g. Foucault's pendulum) to 
    {\it test} the laws of inertia with respect to rotations.---
But the question remains: 
     {\it What physical cause} 
governs the time-evolution of 
the axes of inertial frames and of gyroscopes? 
These three theories 
do not contain such a physical cause, quite to the contrary they
contain an 
     {\it absolute element}, 
namely the non-rotating frame in classical mechanics 
and special relativity,
and the asymptotic non-rotating frame in general relativity  
for isolated systems in asymptotic Minkowski space.
This absolute element is experimentally established 
by Newton's bucket experiment and today by gyroscopes 
or the Sagnac effect in inertial guidance systems.
This absolute element is equivalent (with respect to rotations)
to Newton's absolute space.


\subsection{Mach's Principle}


In the 1870's and 1880's Ernst Mach
   \cite{Mach.Mechanik}
   \cite{Mach.Energy}
stated forcefully the hypothesis 
(as an alternative to Newton's absolute space) 
that the axes of non-rotating frames 
(i.e. axes of gyroscopes) in their time-evolution are
   {\it determined} by, are 
   {\it exactly dragged} by, 
   {\it precisely follow}  
``some average'' of the motion of matter in the universe.
Mach's formulation is given in
     \cite{Mach.Mechanics.quote}.
See also 
     \cite{Ciufolini.Wheeler}.---
Today one generalizes ``matter'' to ``matter-energy'',
     i.e. the principle states that 
     there is exact frame-dragging
     for the axis directions of inertial frames 
     by ``some average'' of the energy currents in the universe. 
This is what we take as the formulation of 
Mach's principle at the level of linear cosmological perturbations. 
At the level of linear perturbations 
gravitational waves do not carry energy 
in the sense of a pseudo-energy-momentum tensor.---
Mach's Principle in this formulation 
follows from general relativity
for linear perturbations of FRW universes with $k =0,$
as will be demonstrated by our result
    Eq.~(\ref{Mach.average.introduction}).


   {\it Many alternative versions}
have been proposed under the name of Mach's principle
by other authors after Mach
    \cite{Barbour.Pfister},
    \cite{refs.Mach-history.Ciufolini}.
We shall discuss Einstein's proposal of 1918
   \cite{Einstein.Mach} 
at the end of sect. 
   \ref{sect.scalar.tensor}.
An incomplete list of ten inequivalent versions
has been given and briefly discussed in
      \cite{Bondi}.
Six of these versions are definitely not satisfied  
for Einstein gravity in a cosmological context, 
and these six versions have 
absolutely no counterpart in the writings of Mach.
Another version is satisfied in Einstein gravity, 
even in a non-cosmological context, but it is too weak: 
     {\it partial dragging} as opposed to 
     {\it exact dragging,} i.e. inertial frames 
     {\it influenced} as opposed to 
     {\it fully determined} by the motion of cosmic matter.
Because of the many conflicting versions
proposed under the name of Mach's principle
by other authors after Mach, 
some physicists have been under the impression 
that the issue is
    ``confusing and ill-defined''.--- 
The version, ``the theory contains no absolute elements,''
explained by J. Ehlers in
     \cite{Barbour.Pfister.Ehlers},
is intimately connected to Mach's 
starting point
(no absolute space)
and formulated in a modern way.
This version is a general requirement, 
but it is not a specific implementation 
in contrast to our result
    Eq.~(\ref{Mach.average.introduction}),
which gives an explicit demonstration 
of how Mach's principle is implemented within general relativity.


\subsection{Gravitomagnetism}


At the time of Mach there was no known mechanism
by which matter in the universe could 
influence the motion of gyroscope axes.
With General Relativity 
came the needed mechanism, gravitomagnetism.
Thirring in 1918
   \cite{Thirring}
considered a rotating infinitely thin spherical shell 
with uniform surface mass density and total mass $M,$ 
and he analyzed the partial dragging of the axes of inertial frames
{\it inside} the rotating shell 
for points near the origin. 
In the weak field approximation,
$G_{\rm N}(M/R)_{\rm shell} \ll 1,$ 
and to first order in the angular velocity
$\Omega_{\rm shell}$
he found that 
the axes of local inertial systems (and gyroscopes) 
at all points near the origin 
rotate 
relative to asymptotic Minkowski space with the 
{\it same} angular velocity 
$\vec{\Omega}_{\rm gyro} = f_{\rm drag} \vec{\Omega}_{\rm shell}.$ 
For the {\it dragging fraction} $f_{\rm drag}$ he obtained 
$f_{\rm drag}= \frac{4}{3} G_{\rm N}(M/R)_{\rm shell} \ll 1.$---
Gravitomagnetic effects in the solar system are exceedingly small.
The Lense-Thirring effect by the rotating earth   
on the spin axes of gyroscopes on Gravity Probe B 
is predicted to be 
only 43 milli-arc-sec per year  
    \cite{GP.B}.---
Brill and Cohen 
     \cite{Brill.Cohen}
and Lindblom and Brill
     \cite{Lindblom.Brill}
analyzed dragging by spherical shells 
to first order in the angular velocity 
but to all orders in $ G_{\rm N}(M/R)_{\rm shell}.$
They found that inside the shell 
one has a Minkowski space which rotates 
relative to asymptotic Minkowski space. 
In the limit of $(M/R)_{\rm shell}$ 
approaching the value for a black hole 
they found perfect dragging 
of inertial axes inside the rotating shell.


Exact dragging 
by the masses in the universe
is required by Mach's principle,
not a small influence.
Therefore one must go to cosmological models.
A simple, cosmologically relevant model is 
a uniform rotational motion of cold matter 
with $\vec{\Omega}_{\rm matter}$ and mass density $\rho$
both constant out to a radius
$R_{\rm rot}$ 
around a gyroscope and $\vec{\Omega}_{\rm matter} = 0 $    
outside $R_{\rm rot}$.
Starting with a weak field perturbation
on a Minkowski background 
this simple model is a 
superposition of a sequence of Thirring shells. 
This gives a dragging fraction which 
     {\it grows quadratically} 
in $R_{\rm rot},$  
$f_{\rm drag} = 2G_{\rm N}(M/R)_{\rm rot}
= (8 \pi /3) G_N \rho R^2_{\rm rot}.$
The Friedmann equation for $k=0$ reads 
$H^2 = (8 \pi /3) G \rho,$ hence $2 G_{\rm N} M_H / R_H = 1,$
where $M_H$ is the mass inside the Hubble radius $R_H.$
This gives 
$f_{\rm drag} = R^2_{\rm rot}/R_{H}^2.$
Although the weak field perturbation on a Minkowski background implies
$f_{\rm drag} = 2G_{\rm N}(M/R)_{\rm rot} \ll 1,$
applying the formula beyond its region of validity
would give $f_{\rm drag} \rightarrow 1$ for 
$R_{\rm rot} \rightarrow R_{H},$ 
a ``satisfying degree of self-consistency'' in the words of
Misner, Thorne, and Wheeler
      \cite{MTW.Mach},
and $f_{\rm drag} \gg 1$ for $R_{\rm rot} \gg R_{H},$ 
the problem of `overdragging'.   
What is needed is an analysis using 
cosmological perturbation theory including superhorizon scales.
This is given in this paper.


\section{Summary and Conclusions}


In this paper we analyze 
realistic cosmological models. 
This is in contrast to toy models involving 
one or many shells (embedded in FRW models), 
infinitely thin and rotating around axes through one point, 
which are contained in
     \cite{C.Klein},
     \cite{Lynden-Bell.1995},
     \cite{Dolezel}.
In models 
     \cite{C.Klein},
     \cite{Dolezel}
``almost all the matter of the universe'' 
must be ``redistributed on the shell''
in order to make frame dragging almost perfect,
i.e. these toy models are extremely far from physical cosmology.
Also we consider realistic cosmological matter of any type
as opposed to the contrived energy-momentum tensors 
discussed in the literature.
We start from a Friedmann-Robertson-Walker 
(FRW) cosmology which is spatially flat (i.e. k = 0). 
We first add the 
most general linear cosmological perturbations
in the vorticity sector  
(vector perturbations), i.e. perturbations derived from 
divergenceless vector fields. 
We ask: Will rotational motions of cosmological matter
exactly drag the axes of any gyroscope 
relative to the directions of geodesics 
from the gyroscope to quasars in the 
asymptotic unperturbed FRW space?
In Sec.~XII we also include scalar and tensor perturbations.


A conference paper about our work appeared in
    \cite{my.gr-qc.02}
and a preliminary version of a full article in
    \cite{my.gr-qc.04}. 
A reaction     
    \cite{Bicak}
to our conference paper 
will be discussed in 
    sect.~\ref{Measured.Matter.Input}.


Our first result is that
the dragging of axes of gyroscopes 
(i.e. axes of local inertial systems)
by matter beyond the $H$-dot radius $R_{dH/dt}$ is 
    {\it exponentially suppressed},
where $R_{dH/dt} \equiv (-dH/dt)^{-1/2},$
$H$ is the Hubble rate, 
and dot is the derivative with respect to cosmic time.
In this sense physics here is decoupled 
from the asymptotic FRW universe.


Our second result refers to the simple model discussed above
(uniform rotation of matter inside a 
radius $R_{\rm rot}$
around a gyroscope). 
It states
that the dragging fraction
$f_{\rm drag}$ approaches the value $1,$ i.e. 
     {\it exact dragging},
exponentially fast as 
$R_{\rm rot}$ gets larger than the H-dot radius.
The dragging fraction as a function of  $R_{\rm rot}$ 
is shown in Fig.~1. 
%
\begin{figure}[h]
\includegraphics{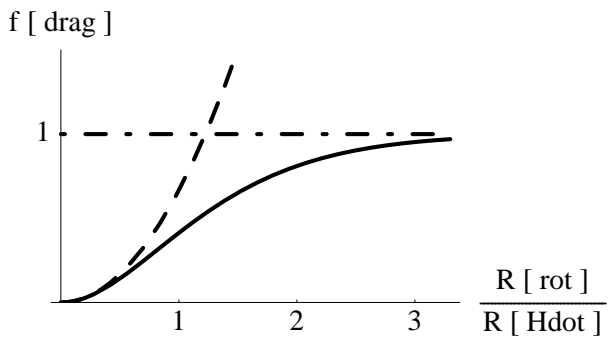}
\caption{\label{caption.text}
The dragging fraction $f_{\rm drag}$ in 
$\vec{\Omega}_{\rm gyro} = 
f_{\rm drag} \, \vec{\Omega}_{\rm matter}$ 
versus $R_{\rm rot}$
for a uniform rotation of cosmological matter  
inside a radius
$R_{\rm rot}$ 
around a gyroscope.
The solid curve shows the result of cosmological
perturbation theory:
As $R_{\rm rot}$ grows beyond the $H$-dot radius, 
$f_{\rm drag}$ approaches the value $1$ exponentially fast,
i.e. 
     {\it exact dragging}. 
The dashed curve is based on extrapolating the results of 
gravitomagnetic perturbation theory on a Minkowski background
beyond the region of validity: 
It shows the 
     {\it problem of `overdragging'},
which is removed by the 
     {\it exponential suppression}
of contributions from super-$H$-dot scales. 
}
\end{figure}
%
The problem of `overdragging' discussed in
   \cite{MTW.Mach}
is removed by the exponential suppression at super-$H$-dot scales, 
which is missing in
   \cite{Lynden-Bell.1995}
and other papers.---
For linear cosmological perturbations both 
$ \Omega_{\rm matter}$   and $\Omega_{\rm gyro}$ 
are infinitesimally small, but their ratio approaches the value $1$
for $R_{\rm rot}$ much larger than the H-dot radius.


Our third result concerns the most general vorticity perturbation in
linear approximation. 
To understand the result one needs two crucial inputs 
from cosmological perturbation theory in the vector sector
(see Sec.~III):
\begin{enumerate}
\item The
   {\it slicing} 
of space-time in slices $\Sigma_{t}$ 
in the vorticity sector is 
   {\it unique},
as emphasized by Bardeen
   \cite{Bardeen}.
\item The 
   {\it intrinsic geometry of 3-space} 
remains 
   {\it unperturbed},
i.e. for vorticity perturbations of FRW with $k = 0$ 
each slice $\Sigma_{t}$ 
remains a 
   {\it Euclidean 3-space} with its global parallelism.
\end{enumerate}
Our result states what specific average 
of the energy flow $\vec{J}_{\epsilon}$
in the universe determines the precession rate 
$\vec{\Omega}_{\rm gyro}$ of gyroscope axes,
\begin{eqnarray} 
&& \vec{\Omega}_{\rm gyro} \,=\, 
- \frac{1}{2} \vec{B}_{\rm g}(r=0) =
\nonumber 
\\
&& 2 G_{\rm N}   \int d^3 
 r^{\prime}             \,  
\{ \frac{1}{r^{\prime}}   [ \vec{r}^{\, \prime} \times 
\vec{J}_{\epsilon} ( \vec{r}^{\, \prime} )] \} \,
\{ \frac{-d}{dr^{\prime}} \, 
\frac{e^{-\mu \, r^{\prime}}}{r^{\prime}} \} ,
\label{Mach.summary}
\end{eqnarray}
where $\vec{r}^{\, \prime}$ is the position of the source 
relative to the gyroscope. 
$\vec{J}_{\epsilon}$ is the energy current, with
$\vec{J}_{\epsilon} = (\rho + p) \vec{v}$ for perfect fluids,
and $\vec{B}_{\rm g}$ is the gravitomagnetic field,
of which the general operational definition is given in 
     section \ref{FIDO.gravitomagn.field}.
All quantities in 
     Eq.~(\ref{Mach.summary})
are 
    {\it directly measurable}, 
i.e. they have a 
    {\it gauge invariant} meaning,
as is explained after
     Eq.~(\ref{B.general}).---
The  weight function is
the density of measured angular momentum of matter 
times $(1/r)$ times the Yukawa force 
$Y_{\mu} (r) = (-d/dr)[(1/r) \exp(-\mu r)].$ 
The exponential suppression is determined by 
$\mu^2 = - 4(dH/dt) = 16 \pi G_{\rm N} (\rho+p).$
Except for the exponential factor the weight function is the same
as in the integrated form of Amp\`ere's law for ordinary magnetism.


Note the 
    {\it fundamental difference}
between cosmological gravitomagnetism,
    Eq.~(\ref{Mach.summary}),
and Amp\`ere's law.
The latter does not hold in a rotating reference frame, 
unless one introduces fictitious forces.
In contrast
    Eq.~(\ref{Mach.summary})
remains valid, as it stands, in a frame 
which is rotating with an angular velocity $\vec{\Omega}^{*}$
relative to asymptotic quasars.
This holds because both sides of
    Eq.~(\ref{Mach.summary})
change by the same amount: On the right-hand side, 
$(\vec{r}^{\, \prime} \times \vec{J}_{\epsilon})$
changes by 
$(\rho + p) [\vec{r}^{\, \prime} \times 
(\vec{\Omega}^{*} \times \vec{r}^{\, \prime})].$
Carrying out the integration gives $\vec{\Omega}^{*}$
with a prefactor 1 for the change on the right-hand side, 
which is equal to the change on the left-hand side.---
The fact that
     Eq.~(\ref{Mach.summary})
holds in any frame which is rotating 
relative to asymptotic quasars 
establishes that the asymptotic inertial frame has 
no influence in cosmological gravitomagnetism;
the time-evolution of local inertial axes is
     {\it determined exclusively}
by the weighted average of cosmological matter flows.


Eq.~(\ref{Mach.summary})
can be rewritten to show explicitely that the precession of 
any gyroscope, $\vec{\Omega}_{\rm gyro}$,
exactly follows, i.e. is exactly dragged
by a weighted average 
of $\vec{\Omega}_{\rm matter}$  
with a weight function $W(r_{PQ}),$
where $P$ and $Q$ are the positions of the gyroscope and the source.
For reasons of symmetry 
under rotations and space reflection,
the velocity field on a shell of a given radius $r_{PQ}$
can only contribute to the gyroscope's precession 
through its term with  
$\{ \ell = 1, \, \mbox{odd parity sequence} \},$
which is equivalent to a rigid rotation  
with the angular velocity $\vec{\Omega}_{\rm matter}(r)$.
We obtain
\begin{eqnarray}
\vec{\Omega}_{\rm gyro} 
&=& \int_0^\infty dr \, \,
\vec{\Omega}_{\rm matter}(r) \, \, W(r),
\nonumber   
\\
W(r) 
&=& \frac{1}{3} \, \mu^2 \, r^3 \, Y_{\mu}(r)
\label{Mach.average.introduction}
\end{eqnarray}
This is our most important result.
It shows that $\vec{\Omega}_{\rm gyro}$ 
is the weighted average  
of $\vec{\Omega}_{\rm matter},$
i.e. the evolution of inertial axes 
     {\it exactly follows}
the weighted average of cosmic matter motion.
The weight function $W(r)$ 
has its normalization (integral over $r$) 
equal to unity, 
as it must be for an averaging weight function in any problem.
That the integral of the weight function 
is equal to unity crucially depends on the 
     {\it exponential cutoff} 
in the Yukawa force $Y_{\mu}(r).$
Details about 
     Eqs.~(\ref{Mach.average.introduction})
are given in the paragraph with
     Eqs.~(\ref{Mach.2}) and
          (\ref{cutoff.function}).---
Our results (valid for 
linear perturbations of a FRW universe 
with $k=0$ and for any equation of state),
particularly    
   Eq.~(\ref{Mach.average.introduction}),
are a clear demonstration 
of how 
{\it Mach's principle is implemented within cosmological general relativity}.


Our aim is to obtain the 
laws of linearized cosmological gravitomagnetism 
in a form which shows the 
correspondence with electromagnetism 
in a $(3+1)$-dimensional split.
In sect. IV  we give the 
general operational definition for the 
gravitoelectric field 
$\vec{E}_{\rm g}$ and the 
gravitomagnetic field $\vec{B}_{\rm g}$
via measurements
by any choice of fiducial observers (FIDOs).
These operational definitions are generally valid, 
i.e. beyond perturbation theory.
The operational definitions need 
     {\it Cartan's formalism} 
with 
     {\it local orthonormal bases} 
(LONBs) and the corresponding fiducial observers.
For our specific problem
we choose FIDOs with 
     {\it spatial basis vectors} 
fixed to directions of 
     {\it geodesics} 
on $\Sigma_t$ from the FIDO to 
     {\it quasars} 
in the asymptotically unperturbed FRW universe.
This construction is clearly independent of the   
chosen coordinate system, 
i.e. it is a 
     {\it gauge-invariant} 
construction.


In sect. V we first give the 
general method to compute 
connection coefficients 
for local orthonormal bases in 
Cartan's formalism.
We apply this general method to 
obtain the connection coefficients for
linear cosmological gravitomagnetism 
and the equations of motion for free-falling test particles.


In sect. VI we compute curvature 
in Cartan's formalism.
The crucial equation for cosmological gravitomagnetism is 
Einstein's $G_{\hat{0} \hat{i}}$-equation
for vorticity perturbations, the 
    {\it momentum constraint}.
It has the form of Amp\`ere's law, except that  
for vorticity perturbations of FRW space there is
an additional term $-4 (dH/dt) \vec{A}_{\rm g},$ 
which is responsible for the exponential suppression 
of super-$H$-dot scales,
\begin{equation}
{\rm curl} \vec{B}_{\rm g} - 4 (dH/dt) \vec{A}_{\rm g} 
= - 16 \pi G_{N} \vec{J}_{\varepsilon}.
\label{mom.constraint.summary}
\end{equation}
In contrast to the Amp\`ere-Maxwell equation 
of ordinary electrodynamics,
the Maxwell term $\partial_t \vec{E}$ is 
    {\it absent} 
in the 
    {\it time-dependent} 
context of gravitomagnetodynamics;
see the comments after 
    Eq.~(\ref{Ampere}).
The solution of 
    Eq.~(\ref{mom.constraint.summary})     
is given in 
    Eq.~(\ref{Mach.summary}).


In sect. VII we give a detailed discussion of  
our three main results for Mach's principle,
which were summarized above.--- 
We then discuss 
Einstein's objection: 
``Mach conjectured that inertia would have to 
depend upon the interaction of masses ...  
and their interactions as the original concepts. 
The attempt at such a solution 
does not fit into a consistent field theory.''
We point out the remarkable fact that 
the solution of Einstein's 
     {\it local field equation} 
for $G_{\hat{0} \hat{i}},$
i.e. of the momentum constraint 
    equation~(\ref{mom.constraint.summary}),
has the form of an 
    {\it instantaneous action-at-a-distance,}
    Eq.~(\ref{Mach.summary}). 
See also refs.
    \cite{Lindblom.Brill},
    \cite{instantaneous.inertial.frame}.


In sect. VIII  we explain that the density of
     {\it measured angular momentum,} 
formed from the LONB components 
$T^{\hat{0}}_{\, \, \hat{i}},$ 
is the relevant input for Mach's principle
on the right-hand side of Einstein's equation, and 
     {\it not} 
the density of conserved angular momentum, 
e.g. the coordinate-basis component $T^0_{\, \phi},$ 
which was considered by other authors, e.g.
     \cite{Lynden-Bell.1995}. 
This is the reason why these and other authors 
did 
     {\it not} 
obtain the Yukawa suppression 
beyond the $H$-dot radius. 
This Yukawa suppression is crucial 
for obtaining the normalization to unity 
in the weight function $W(r)$ in 
     Eq.~(\ref{Mach.average.introduction}),
and it is crucial to 
remove the problem of `overdragging'.


In sect. IX we discuss Einstein's objection 
``If you have a tensor
$T_{\mu \nu}$ and not a metric, then this does not meaningfully 
describe matter. .... The statement that matter
by itself determines the metric ... is meaningless.'' 
From this objection Einstein drew the conclusion that
``one should no longer speak of Mach's principle.''--- 
His objection applies to 
    {\it coordinate-basis components} 
$T_{\mu \nu},$ but we explain why it does not apply to 
    {\it LONB components} 
$T_{\hat{a} \hat{b}}.$
LONB components are determined directly 
by matter measurements using only 
$\eta_{\hat{a} \hat{b}} = {\rm diag} \{ -1,+1,+1,+1  \},$ 
which is available 
    {\it before} 
having solved Einstein's equations.
In contrast $T_{\mu \nu}$ needs matter measurements plus $g_{\mu \nu},$ 
which is available only 
    {\it after} 
having solved Einstein's equations.


In sect. X we explain that 
the vanishing of 
    {\it local vorticity} 
measured by 
non-rotating observers is 
    {\it not relevant} as a test of Mach's principle, 
because Mach's principle requires an 
    {\it average} 
over all matter in the universe with a given weight function.
Therefore we conclude that the claim 
by Ozsv\`ath and Sch\"ucking 
that their Bianchi IX model violates Mach's principle
is not conclusive.


In sect. XI  we derive the 
      {\it boundary conditions} 
needed when solving Einstein's equations 
for a system in 
      {\it asymptotic Minkowski geometry}, 
if one does not restrict the allowed coordinate systems
to those which are asymptotically non-rotating.
The boundary conditions are needed to 
encode the effects of sources outside the system considered, 
the effects of cosmological sources.---
For linear vorticity perturbations of FRW cosmologies 
no boundary conditions of the encoding type are needed
because of the exponential suppression of super-$H$-dot scales.
This result is in 
contrast to Einstein's conclusion 
that the problem of inertia and 
of boundary conditions 
could only be solved if the universe is spatially closed.


In sect. XII we add 
   {\it scalar and tensor perturbations},
and give the mathematical expression 
for Mach's principle in this more general case.---
We  discuss 
Einstein's formulation 
of Mach's principle in 1918
       \cite{Einstein.Mach}.
For linear cosmological perturbations 
Einstein's formulation is 
(1)~too strong to be valid, as has been stated many times
(gravitational waves can exist without matter), and 
(2)~unnecessarily strong, as we shall explain, 
for Mach's original purpose
(the dragging of axis directions of inertial frames). 


In sect. XIII we list 
open problems 
in connection with Mach's principle.


In a companion paper 
      \cite{my.FRW.not.spatially.flat}
we shall present the results, 
analogous to those in this paper, 
for perturbations of FRW universes with $k= \pm 1,$
i.e. spatially spherical or hyperbolic.---
We use the conventions of Misner, Thorne, and Wheeler
   \cite{MTW}.


\section{Vorticity Perturbations}


Linear cosmological perturbations, 
see J. Bardeen  
   \cite{Bardeen}, 
Kodama and Sazaki
   \cite{Kodama.Sazaki},
decouple into three sectors,
3-scalars (density perturbations), 
3-vectors (vorticity perturbations), and 
3-tensors (gravitational waves).
In the vector sector all quantities must be constructed from
3-vector fields with vanishing divergence.


There are two crucial results, 
which make the analysis of vector perturbations
very simple:
\begin{enumerate}
\item 
There is 
     {\it no ambiguity}
about the 
     {\it slicing} 
of space-time 
in the vector sector: 
Since the 3-scalar $\delta g_{00}$ must be zero in the vector sector,
the lapse function is unperturbed.
This was already emphasized by Bardeen 
   \cite{Bardeen}:
``The gauge-transformation properties 
of vector perturbations are much simpler
than for scalar perturbations,
since now there is 
no gauge ambiguity about the time coordinate.'' 
\item
The 
    {\it intrinsic geometry} 
of each 
$\Sigma_{t}$ remains 
    {\it unperturbed}
for linear vector perturbations.
To prove this,
one starts from the general metric in the vector sector,
and one makes a gauge transformation 
    \cite{Bardeen}
such that the $3$-metric is unperturbed.---
Hence for vector perturbations of FRW with $k = \{ 0,+1,-1 \}$
the fixed-time slices $\Sigma_t$ remain 
Euclidean 3-spaces $E^3,$
resp 3-spheres $S^3,$ resp hyperbolic 3-spaces $H^3$.
\end{enumerate}


Since linear vector perturbations leave 
3-space in FRW unperturbed, 
the natural choices of spatial coordinates are 
comoving Cartesian coordinates 
for Euclidean 3-space in FRW with $k = 0,$ resp 
comoving spherical coordinates 
for 3-space in FRW with $k= \{ 0, \pm 1 \}$.
The line elements are
\begin{equation}
ds^{2}=-dt^{2}+
\sum_i [(ah_{i})^{2}(dx^{i})^{2}+
 2 \beta_{i} dx^{i}dt],
\label{line.element} 
\end{equation}
where $\beta^{i}$ is the shift 3-vector, 
the only quantity referring to vorticity perturbations
in our coordinate choice, and $\beta_i = (ah_i)^2 \beta^i.$
For a FRW background with $k = 0$ and for $\Sigma_t$ 
in comoving Cartesian coordinates
\begin{equation}
h_i = 1,  \quad i = 1, 2, 3.
\end{equation}
For a FRW background with $k = \{0, +1, -1 \}$ and for $\Sigma_t$ 
in comoving spherical coordinates
$\{ \chi, \theta, \phi  \}$
\begin{eqnarray}
h_{\chi}=1, \quad h_{\theta} &=& R(\chi), \quad h_{\phi}=R(\chi) \sin \theta,
\nonumber
\\
R(\chi)  &=&  \{ \chi, \, \sin \chi, \, \sinh \chi  \}.
\end{eqnarray}
In sections III, IV, V 
we give 
    {\it unified computations for FRW with} $k= \{0, \pm 1 \},$ 
but afterwards in this paper we specialize to $k = 0.$


For perturbed FRW with $k=0$ all $\Sigma_t$ 
retain a 
Euclidean 3-geometry.
With our choice of comoving 
Cartesian coordinates for Euclidean 3-space
the 3-coordinate lines are 
     {\it geodesics} 
in 3-space. 
In all other gauges,
where a perturbed 3-metric $g_{\mu \nu}$ is used for Euclidean 3-space
(in Bardeen's notation with $H_{T}^{(1)} \neq 0$),
the 3-coordinate lines 
will quickly wind up more and more
into spirals 
relative to geodesics on $\Sigma_t$.
Hence all other gauges (particularly time-orthogonal) 
give a very awkward way 
to coordinatize Euclidean 3-space.


We consider vector perturbations in an 
     {\it asymptotically unperturbed FRW universe},
i.e. $\beta^{i} \rightarrow 0$ for $r \rightarrow \infty.$
We fix our coordinates to 
quasars in the asymptotic unperturbed FRW space.
(This is analogous to ``star-fixed coordinates'' 
for the physics of rotating black holes.)
This 
     {\it fixes our gauge completely}.---
The basis vectors in the coordinate basis,
$\bar{e}_i(P) = \partial / \partial x^{i},$ 
point along geodesics in $\Sigma_{t}$ from $P$
to fixed quasars in the asymptotic FRW space.
The case of perturbed FRW with $k = 0$ is particularly simple, because
$3-$space remains Euclidean everywhere, 
hence each $\Sigma_t$ possesses global parallelism.


It will turn out in
    Eq.~(\ref{shift.vectorpotential})
that our shift vector 
$\vec{\beta}$
is equal to the gravito-magnetic vector potential 
$\vec{A}_{{\rm g}},$
\begin{equation}
\vec{\beta} = \vec{A}_{\rm g}.
\label{beta.A}
\end{equation}
Our sign convention for the shift vector $\beta^i$ is dictated
by our wish to have no minus sign in 
    Eq.~(\ref{beta.A}).
Our sign convention for the shift agrees with Misner, Thorne, and Wheeler
    \cite{MTW},
and it is the opposite of Bardeen's
    \cite{Bardeen}.---
With our gauge-fixing 
the gravito-magnetic vector potential $\vec{A}_{{\rm g}}$
is directly related to a 
    {\it measurable geometric quantity}: 
div  $\vec{A}_{\rm g} =  0,$ and
curl $\vec{A}_{\rm g} = -2 \, \vec{\Omega}_{\rm gyro},$
where $\vec{\Omega}_{\rm gyro}$ is the precession 
of local gyroscopes
relative to the geodesics to asymptotic quasars.
See
    Eqs.~(\ref{var.op.def.B}) and 
         (\ref{fields.from.vector.potential}) 
below.


   {\it Bardeen's gauge-invariant combination}
$\Psi$ is defined by
$\Psi \equiv B^{(1)} - \frac{1}{k} \dot{H}_T^{(1)}$ 
(in Bardeen's notation).
$B^{(1)}$ is the shift amplitude, 
and $H_T^{(1)}$ is the amplitude of the perturbation of the 
   {\it 3-metric} 
chosen for the unperturbed 
   {\it 3-geometry}.
Evidently the gauge-invariant 
amplitude $\Psi$ is equal to the
shift amplitude $B^{(1)}$ in the 
gauge which is fixed by 
$H^{(1)} \equiv 0,$
i.e. in the gauge, where $^{(3)}g_{ij}$ is unperturbed.
Bardeen emphasized in
     \cite{Bardeen.1988}:
``Much has been made by some authors of the advantages 
of using gauge-invariant variables. However ...''
We now transcribe Bardeen's formulation
     \cite{Bardeen.1988} 
from the scalar sector to the vector sector:
``The advantages of the gauge-invariant variable $\Psi$
are the advantages of working with the shift amplitude 
$B^{(1)}$ in the gauge 
where $^{(3)}g_{ij}$ is unperturbed,
no more and no less.''


The fact stated in the preceeding sentence 
expressed in our notation: 
The 
    {\it shift} 
vector $\beta^i$ 
in the 
    {\it Cartesian gauge} 
(for vector perturbations of FRW with $k=0$) is
    {\it equal} 
to 
    {\it Bardeen's gauge-invariant variable} 
$\Psi$ times his vector harmonic $Q^{(1)i} (x)$
(for a given wave number and polarization)
apart from the sign discussed above,
\begin{equation}
\beta^i (x,t) 
= A^i_{\rm g}
= - \Psi (t) Q^{(1)i} (x).
\label{beta.equiv.Psi}
\end{equation}
{\it Our gravito-magnetic vector potential} 
$\vec{A}_{\rm g}$ 
is directly proportional to Bardeen's  
     {\it gauge-invariant combination} $\Psi.$
Working with 
     {\it our gauge}, i.e. with comoving
Cartesian (or spherical) coordinates
for Euclidean 3-space, is totally 
    {\it equivalent} 
to working with 
    {\it Bardeen's gauge-invariant potential} $\Psi.$


\section{Fiducial Observers and Gravitomagnetic Field}        
\label{FIDO.gravitomagn.field}


Our aim is to obtain the laws of linearized gravito-magnetism 
in a formulation analogous to electromagnetism 
with a split to $(3+1)$-dimensions. 
What is the operational definition for the 
    {\it gravitoelectric field} $\vec{E}_{\rm g}$
and 
    {\it gravitomagnetic field} $\vec{B}_{\rm g}$?
According to the equivalence principle 
for a free-falling, non-rotating observer there are 
no gravitational forces at his position,
$\vec{E}_{\rm g}=0,\  
 \vec{B}_{\rm g}=0.$
It all depends on the choice of 
    {\it fiducial observers} 
(FIDOs) with their 
    {\it local ortho-normal bases} 
(LONBs),
see Thorne et al 
    \cite{Thorne}.
For FIDOs which are not free-falling and/or which are rotating
(relative to local gyroscope axes) there are gravitational forces 
(inertial forces, fictitious forces, e.g. the Coriolis force).


The 
    {\it operational definitions} of 
$\vec{E}_{\rm g}$ and 
$\vec{B}_{\rm g}$ are independent of perturbation theory.
They involve a family of world lines of fiducial observers, FIDOs, 
of any given choice,
such that through each space-time point P 
there passes precisely one world line,
i.e. the family of world lines forms a congruence.
The measurements of a FIDO are directly encoded by the 
components (of momentum $\bar{p}$ and of gyroscope spin $\bar{S}$)
with respect to the FIDO's local orthonormal basis,
where $\bar{e}_{\hat{0}}(P) = \bar{u}_{\rm FIDO}(P),$ 
and $\bar{e}_{\hat{i}}(P)$ are given by any definite choice.
Hats always refer to LONBs, and bars designate space-time vectors.---
To convert local measurements 
by a FIDO to LONB-components 
$p^{\hat{a}}$ and $S^{\hat{a}}$
one only needs 
$\eta_{ \hat{a} \hat{b} } = {\rm diag} \{-1,+1, +1, +1 \},$
but one does not need $g_{\mu \nu},$ 
which is not available before 
having solved Einstein's equations. 
Therefore it is 
    {\it necessary to abandon} 
vector components in a 
    {\it coordinate basis}, 
to work with LONBs, 
and to use the formalism of {\it  \'{E}lie Cartan} 
   \cite{MTW},
   \cite{Cartan}.---
Measuring $\{ \vec{E}_{\rm g}, \vec{B}_{\rm g} \}$
involves FIDOs measuring first time-derivatives along their world lines, 
on the one hand of the momentum components $p_{\hat{i}}$ 
of free-falling quasistatic test particles, 
and on the other hand of the spin components $S_{\hat{i}}$ of gyroscopes 
carried along by the FIDOs,
\begin{eqnarray}
\frac{d}{dt} p_{\hat{i}} \equiv
m E_{\hat{i}}^{\rm g} \quad \mbox{free-falling quasistatic 
test particle,}
\label{op.def.E}
\\
\frac{d}{dt} S_{\hat{i}} \equiv
- \frac{1}{2} [\vec{B}_{\rm g} \times \vec{S}]_{\hat{i}}
\quad \mbox{gyro comoving with FIDO,}
\label{op.def.B} 
\end{eqnarray}
where $t$ is the local time measured by the FIDO.
Arrows denote 3-vectors in the tangent spaces spanned by 
the spatial legs of LONBs.
$\vec{E}_{\rm g} \equiv \vec{g}$ is the gravitational acceleration
of free-falling quasistatic test particles relative to the FIDO.---
The operational definitions of 
    Eqs.~(\ref{op.def.E},\ref{op.def.B}) 
are the same as for a classical charged spinning test particle 
in an electromagnetic field 
except that the charge $q$ is replaced by $m,$
and the gyromagnetic ratio $q/(2m)$ is replaced by $1/2.$
      Eq.~(\ref{op.def.B}) 
gives the angular velocity of precession of the 
gyroscope's spin axis relative to the axes of the FIDO,
\begin{equation}
\vec{\Omega}_{\rm gyro} \equiv -\frac{1}{2} \vec{B}_{\rm g},
\label{var.op.def.B}
\end{equation}
which is an equivalent operational definition of 
$\vec{B}_{\rm g}.$---
For gyroscopes carried along by FIDOs the magnitude 
of $\vec{S}$ is constant, 
and for non-rotating FIDOs 
the components of the vector $\vec{S}$ are constant,
because a gyroscope experiences no mechanical torques by construction
and no gravitational torques by the equivalence principle.


From the operational definitions of 
$\{\vec{E}_{\rm g},   
   \vec{B}_{\rm g} \}$  it follows immediately
that these fields are identical (apart from an overall sign)
to the connection coefficients 
$(\omega_{ \hat{a} \hat{b} })_{\hat{0}}$  
for a displacement along the FIDO's world line.---         
The  connection 1-forms 
$ \tilde{\omega}^{\hat{a}}_{\ \hat{b}}$ 
resp. their components in LONBs (= Ricci rotation coefficients),
$(\omega^{\hat{a}}_{\ \hat{b}})_{\hat{c}},$  
are defined by
\begin{equation}
\nabla_{\hat{a}} \bar{e}_{\hat{b}} = \bar{e}_{\hat{c}}
(\omega^{\hat{c}}_{\ \hat{b}})_{\hat{a}}.
\end {equation}
In words: the Ricci rotation coefficients 
$(\omega^{\hat{c}}_{\ \hat{b}})_{\hat{a}}$
give the rotation resp. the Lorentz boost 
$(\omega^{\hat{c}}_{\ \hat{b}})$
of the LONBs under 
parallel transport along $\bar{e}_{\hat{a}}.$
Parallel transport of $\bar{u} = \bar{e}_{\hat{0}}$ 
along $\bar{u}$ is given by free fall, 
parallel transport for $\bar{e}_{\hat{i}}$
is given by gyroscope axes.
Relative to FIDOs the equations of motion for free-falling test particles  
(geodesic equation) and for spin axes of gyroscopes
(Fermi transport) specialized to gyroscopes 
carried along by a FIDO are
\begin{equation}
\frac{dp^{\hat{a}}}{dt} +
(\omega^{\hat{a}}_{\ \hat{b}})_{\hat{c}} \,
p^{\hat{b}} \frac{dx^{\hat{c}}}{dt}=0, \quad
\frac{dS^{\hat{i}}}{dt} + 
(\omega^{\hat{i}}_{\ \hat{j}})_{\hat{0}} \, S^{\hat{j}}=0.
\label{geodesic.equation.and.Fermi.transport}
\end{equation}
With Eqs.~(\ref{geodesic.equation.and.Fermi.transport})
the operational definitions 
     Eqs.~(\ref{op.def.E}, \ref{var.op.def.B})
are translated into the equivalent definitions 
involving connection coefficients with a 
displacement index $\hat{0}$,
namely a Lorentz boost $\omega_{\hat{i} \hat{0}}$ 
per unit time (acceleration $\vec{g}=\vec{E}_{\rm g}$) resp 
a rotation angle $\omega_{\hat{i} \hat{j}}$
per unit time (angular velocity $\vec{\Omega}_{\rm gyro}=
-\frac{1}{2}\vec{B}_{\rm g}$),
\begin{equation}
(\omega_{\hat{i}\hat{0}})_{\hat{0}} \equiv
- E_{\hat{i}}^{\rm g}, \quad 
(\omega_{\hat{i}\hat{j}})_{\hat{0}} \equiv
- \frac{1}{2}  B_{\hat{i}\hat{j}}^{\rm g},
\label{2nd.op.def.EB}
\end{equation}
where $B_{\hat{i}\hat{j}} \equiv 
\varepsilon_{\hat{i}\hat{j}\hat{k}}B_{\hat{k}}, \,
\Omega_{\hat{i}\hat{j}} \equiv 
\varepsilon_{\hat{i}\hat{j}\hat{k}} \Omega_{\hat{k}},$
and
$\Omega_{\hat{i}\hat{j}} = (\omega_{\hat{i}\hat{j}})_{\hat{0}}.$
The last equation (angular velocity) 
gives the motivation for using the letter $\omega$
for Ricci rotation coefficients (connection coefficients).
Only displacments along $\bar{e}_{\hat{0}}$ appear in 
    Eq.~(\ref{2nd.op.def.EB}),
because for quasistatic test particles
(typically initially at rest at the position of the FIDO) 
the spatial separation from the FIDO 
grows only quadratically in time.


      {\it Our choice of FIDOs}: 
The world lines of our FIDOs are at fixed
$x^{i}$ in our coordinates, which are fixed to quasars
in the asymptotic FRW universe.
The 3-velocity of our FIDOs 
relative to the normals on
$\Sigma_t$ is equal to the shift 3-vector 
$\beta^{i}$ in 
     Eq.~(\ref{line.element}).---
The construction of the world-lines of FIDOs is equivalent 
to the gauge-invariant statement that our FIDOs are 
     {\it at rest in the Hubble frame}  
given by asymptotic quasars on $\Sigma_t.$--- 
We choose the 
     {\it spatial basis vectors} 
of our FIDOs, $\bar{e}_{\hat{i}}(P)$,
fixed to directions of 
     {\it geodesics} 
on $\Sigma_t$ from $P$ to 
     {\it quasars} 
in the asymptotic FRW universe, again a 
     {\it gauge-invariant} 
statement. 
Hence we fix $\bar{e}_{\hat{i}}(P)$ 
in the same spatial directions as 
$\bar{e}_{i}(P) \equiv \partial/\partial x^{i},$
i.e. in 4-space the directions of $\bar{e}_{\hat{i}}$ and
$\bar{e}_{i}$ differ by a pure Lorentz boost, see 
   Eq.~(\ref{expansion of basis vectors}) below.


\section{Connection Coefficients and            
  equations of motion for matter in gravitomagnetism}


Cartan's formalism with 
local orthonormal tetrads is crucial 
for gravitomagnetism, because
(1) the operational definitions
of the gravito-electric and gravito-magnetic fields
(independent of perturbation theory) 
are given by measurements of the corresponding FIDOs,
    Eqs.~(\ref{op.def.E}) - 
         (\ref{var.op.def.B}),
and 
(2) the measured LONB-components of the 
gravito-electric and gravito-magnetic fields
are identical to connection coefficients 
in the local orthonormal basis of the FIDO,
    Eqs.~(\ref{2nd.op.def.EB}). 
A third reason, 
why it is necessary to work in Cartan's formalism with
local orthonormal tetrads, will be presented in section IX.


The {\it computation of connection coefficients}
in Cartan's formalism: We give a short, self-contained
derivation of the formulae needed 
from Cartan's formalism with LONB-tetrads.
This should be helpful for readers only familiar 
with the literature on cosmological perturbation theory 
and with standard one-year graduate courses on General Relativity.
Readers not interested in calculational methods 
can go directly to the results starting with 
    Eq.~(\ref{connection.coeff.Mink.background}).


\subsection{Cartan's first equation in LONB components}


The 
     {\it first step} 
is to express 
our choice of LONBs $\bar{e}_{\hat{a}}(P),$
given at the end of the previous section, 
in terms of the coordinate bases 
$\bar{e}_{\alpha}(P)=\partial / \partial x^{\alpha}  ,$ i.e. 
$\bar{e}_{\hat{a}}=(e_{\hat{a}})^{\alpha}\bar{e}_{\alpha}.$
To first order in 
$(\beta^{i}/c)$
\begin{equation}
\bar{e}_{\hat{0}}=\bar{e}_{0}, 
\qquad \bar{e}_{\hat{k}}=
\frac{1}{ah_{k}}(\bar{e}_{k}+\beta_{k}\bar{e}_{0}).
\label{expansion of basis vectors}
\end{equation}
Since the spatial LONBs point in the 
same spatial directions as the spatial
coordinate bases, we use latin letters 
from the middle of the alphabet both
for spatial LONBs (with hat) and for 
spatial coordinate bases (without hat).
The dual bases (basis 1-forms) 
$\tilde{\theta}^{\hat{a}}$ for LONBs 
are defined by 
$\langle \tilde{\theta}^{\hat{a}}, \bar{e}_{\hat{b}} \rangle =
\delta^{\hat{a}}_{\hat{b}},$ 
where tildes designate space-time 1-forms.
The LONB 1-forms $\tilde{\theta}^{\hat{a}}$ are expanded in the  
coordinate basis 1-forms, $\tilde{\theta}^{\alpha} = 
\tilde{d}x^{\alpha}$, i.e.                       
$\tilde{\theta}^{\hat{a}}= 
(\theta^{\hat{a}})_{\alpha}\tilde{\theta}^{\alpha},$ by
\begin{equation}
\tilde{\theta}^{\hat{0}}=
\tilde{\theta}^{0}-\beta_{k} \tilde{\theta}^{k},\qquad 
\tilde{\theta}^{\hat{k}}=ah_{k}  \tilde{\theta}^{k}.
\label{expansion of basis 1-forms}
\end{equation} 
The coefficients of the inverse expansion, 
i.e. coordinate bases 
in terms of LONBs, are
$(e_{\alpha})^{\hat{a}} = (\theta^{\hat{a}})_{\alpha}$ resp.
$(\theta^{\alpha})_{\hat{a}} = (e_{\hat{a}})^{\alpha}.$


The 
     {\it second step} 
is computing the exterior derivative $d$ of 
the basis-1-forms, $(d\theta^{\hat{c}})_{\alpha \beta} \equiv
\partial_{\alpha}(\theta^{\hat{c}})_{\beta} -
\partial_{\beta} (\theta^{\hat{c}})_{\alpha},$
where $[\alpha \beta]$ must be in the coordinate basis.
Then one converts to components $[\hat{a} \hat{b}]$ in the LONB,
\begin{equation}
(d\theta^{\hat{c}})_{\hat{a} \hat{b}} \equiv
- C_{\hat{a} \hat{b}}^{\ \ \hat{c}} =
(e_{\hat{a}})^{\alpha} 
[\partial_{\alpha} (\theta^{\hat{c}})_{\beta}
-\partial_{\beta}  (\theta^{\hat{c}})_{\alpha}]
(e_{\hat{b}})^{\beta}.
\label{exterior.derivative.in.LONB}
\end{equation} 
The coefficients $C_{\hat{a} \hat{b}}^{\ \ \hat{c}}$
are identical to the {\it commutation coefficients} of the
basis vectors, 
\begin{equation}
[\bar{e}_{\hat{a}},\bar{e}_{\hat{b}}] \equiv
C_{\hat{a}\hat{b}}^{\ \ \hat{c}} \bar{e}_{\hat{c}}.
\end{equation}
This is easily shown by using 
$\partial_{\alpha} \langle \tilde{\theta}^{\hat{c}}, 
\bar{e}_{\hat{b}} \rangle = 0
= [\partial_{\alpha} 
(\theta^{\hat{c}})_{\beta}] (e_{\hat{b}})^{\beta} +    
(\theta^{\hat{c}})_{\beta} [\partial_{\alpha}(e_{\hat{b}})^{\beta}].$


The 
     {\it third step} 
is obtaining the connection coefficients 
from the commutation coefficients.
The definition of the connection via basis 1-forms is
$(\nabla_{\alpha} \theta^{\hat{c}})_{\beta}
=-(\omega^{\hat{c}}_{\ \hat{d}})_{\alpha}
(\theta^{\hat{d}})_{\beta}.$
We take both the displacement index $\alpha$ and the equation's 
component index $\beta$ in the coordinate basis,
and we antisymmetrize in $[\alpha \beta].$ This makes 
the Christoffel symbols $\Gamma^{\rho}_{\ \alpha \beta}$ 
on the left-hand side disappear,
since they are symmetric in $\alpha,\beta.$
Hence the left-hand side reduces to 
$(d \theta ^{\hat{c}})_{\alpha \beta}$.
Dropping the equation's component indices 
$ [\alpha,\beta]$ gives
\begin{equation}
\tilde{d} \tilde{\theta}^{\hat{c}}
= - \tilde{\omega}^{\hat{c}}_{\ \hat{d}}
\wedge \tilde{\theta}^{\hat{d}}.
\label{first.Cartan}
\end{equation}
This is 
     {\it Cartan's first equation}.
The wedge product (exterior product) of two 1-forms 
$\tilde{\sigma}$ and $\tilde{\rho}$ is
$(\sigma \wedge \rho)_{\alpha\beta}
\equiv 
\sigma_{\alpha}\rho_{\beta}-\sigma_{\beta}\rho_{\alpha}.$
Taking Cartan's first equation in LONB components
gives the commutation coefficients, 
    Eq.~(\ref{exterior.derivative.in.LONB}), 
on the left-hand side,
and the right-hand side simplifies in LONB 
because $(\theta^{\hat{d}})_{\hat{b}}=
\delta^{\hat{d}}_{\hat{b}}.$ 
Hence 
     {\it Cartan's first equation in LONB components} is
\begin{equation}
C_{\hat{a}\hat{b}}^{\ \ \hat{c}}=
(\omega^{\hat{c}}_{\ \hat{b}})_{\hat{a}} -
(\omega^{\hat{c}}_{\ \hat{a}})_{\hat{b}}.  
\end{equation}
This equation is easily solved for the rotation coefficients,
\begin{equation}
(\omega_{\hat{c}\hat{b}})_{\hat{a}}= \frac{1}{2}[ \,
C_{\hat{c}\hat{b}\hat{a}} +
C_{\hat{c}\hat{a}\hat{b}} -
C_{\hat{b}\hat{a}\hat{c}} \, ].
\label{connection.coeff.commutation.coeff} 
\end{equation}


\subsection{Connection coefficients for vorticity perturbations 
on a Minkowski background}

  
To first order in $\beta^{i}$
and with Cartesian spatial coordinates on $\Sigma_{t},$
the commutation coefficients 
are very simple to compute,
because only $(\theta^{\hat{0}})_{i}=-\beta_{i}$ 
is space-time dependent, and the prefactors in 
   Eq.~(\ref{exterior.derivative.in.LONB}), 
$(e_{\hat{a}})^{\alpha}$ and $(e_{\hat{b}})^{\beta},$
can be set to $1$. 
Hence $C_{\hat{a}\hat{b}}^{\ \ \hat{0}}=
(d\beta)_{ab},$ and
\begin{eqnarray}
    (\omega_{\hat{i}\hat{0}})_{\hat{0}} 
&\equiv& \, - \quad \, E_{\hat{i}}^{\rm g} \, \, 
  = \quad  \, \, \, \partial_{t}\beta_{i}, \nonumber \\
    (\omega_{\hat{i}\hat{j}})_{\hat{0}} 
&\equiv& \, - \frac{1}{2} \, \,
B_{\hat{i} \hat{j}}^{\rm g} \, \,
  = \, \, - \frac{1}{2} (d \beta)_{ij} \, \,
  = \, \, (\omega_{\hat{i}\hat{0}})_{\hat{j}}, \nonumber \\
    (\omega_{\hat{i}\hat{j}})_{\hat{k}} 
&=& 0.
\label{connection.coeff.Mink.background}
\end{eqnarray} 
All connection coefficients 
with respect to LONB's are 
     {\it directly measurable} 
(in contrast to Christoffel symbols, which refer to
coordinate bases).


\subsection{Connection coefficients for vorticity perturbations  
on FRW with $\{ k=0, \pm 1 \}$}


It is again straightforward to compute 
the commutation coefficients and the connection coefficients
using 
   Eqs.~(\ref{expansion of basis vectors},
         \ref{expansion of basis 1-forms},
         \ref{exterior.derivative.in.LONB}, 
         \ref{connection.coeff.commutation.coeff}),
\begin{eqnarray}
(\omega_{\hat{i}\hat{0}})_{\hat{0}}&\equiv&
- E_{\hat{i}}^{\rm g}=
\frac{1}{a} \partial_{t}(a \beta_{\hat{i}}),
\label{dbeta/dt}
\\
(\omega_{\hat{i}\hat{j}})_{\hat{0}} &\equiv&
-\frac{1}{2}
B_{\hat{i} \hat{j}}^{\rm g}=
-\frac{1}{2}(d\beta)_{\hat{i} \hat{j}},
\label{curl beta}
\\
(\omega_{\hat{i}\hat{0}})_{\hat{j}} &=&
-\frac{1}{2}
B_{\hat{i} \hat{j}}^{\rm g}
+\delta_{\hat{i}\hat{j}}H,
\label{omega.i0j}
\\
(\omega_{\hat{i}\hat{j}})_{\hat{k}}
&=& 
\delta_{\hat{i}\hat{k}} (H \beta_{\hat{j}}+
\partial_{\hat{j}} L_{\hat{i}})-
\delta_{\hat{j}\hat{k}} (H\beta_{\hat{i}}+
\partial_{\hat{i}} L_{\hat{j}}), \quad
\label{omega.ijk}
\end{eqnarray}
where $L_{\hat{i}} \equiv \log h_{i}, \, \,
\partial_{\hat{i}} = (a h_i)^{-1} \partial_i, \, \, $ 
and $H$ is the Hubble rate.
In 
    Eq.~(\ref{omega.ijk})
there is no summation over repeated indices 
on the right-hand side.---
The Hubble term in
    Eq.~(\ref{omega.i0j})
follows directly by comparing the 
operational definitions of
the Hubble rate and of the connection coefficients.


Since we work to first order in the vorticity 
perturbations (i.e. in $\beta ^{i}$), we can identify 
$\vec{E}_{\rm g}$ and 
$\vec{B}_{\rm g}$ 
with vectors in the tangent spaces to $\Sigma_{t}.$        
From Eqs.~(\ref{dbeta/dt}, \ref{curl beta}), 
we see that the shift vector $\vec{\beta}$
must be identified with the gravitomagnetic vector potential
$\vec{A}_{\rm g}$.  
From 
    Eqs.~(\ref{line.element}, \ref{dbeta/dt}, \ref{curl beta}) 
follow
\begin{eqnarray}
&& \quad \quad \quad \quad g_{0i} \, = \, \beta_i \, = \, (A_{\rm g})_i
\label{shift.vectorpotential} 
\\
&&\vec{B}_{\rm g}={\rm curl} \vec{A}_{\rm g}, \quad 
\vec{E}_{\rm g}=-\frac{1}{a}\partial_{t}(a\vec{A}_{\rm g}), \quad
\label{fields.from.vector.potential}
\\
&& \quad \quad {\rm curl} \vec{E}_{\rm g}+
\frac{1}{a^{2}}\partial_{t}(a^{2}\vec{B}_{\rm g})=0.
\label{Faraday}
\end{eqnarray}
  Equations~(\ref{fields.from.vector.potential})~and(\ref{Faraday}) 
are identical with the homogeneous equations for 
electromagetism in FRW space-times with $k=0, \pm 1$.


\subsection{Equation of motion for free-falling test particles}


The equation of motion (geodesic equation) for free-falling 
test particles of arbitrary velocities $v \leq c$ 
(e.g. photons)  
in linear vorticity perturbations on a 
     {\it Minkowski background} 
reads
\begin{equation}
\frac{d}{dt}(p_{\hat{i}}) =
\varepsilon [\vec{E}_{\rm g}+
(\vec{v} \times \vec{B}_{\rm g})]_{\hat{i}}.
\label{Lorentz.law}
\end{equation}
This is identical with the Lorentz law for electromagnetism, 
except that the charge $q$ 
is replaced by the energy $\varepsilon$ of the test particle,
$\varepsilon = m_0 (1 - \beta^2)^{-1/2}$ for massive particles,
and $\varepsilon = h \nu$ for photons.
With Eq.~(\ref{var.op.def.B}) 
and in a stationary gravitomagnetic field ($\vec{E}_{\rm g}=0$), 
Eq.~(\ref{Lorentz.law}) becomes 
$\frac{d}{dt}(p_{\hat{i}}) = 
- 2 \varepsilon [\vec{v} \wedge \vec{\Omega}_{\rm gyro}]_{\hat{i}},$
the Coriolis force law. 
Note that $\Omega_{\rm gyro}$ is minus the
rotation velocity of the FIDO relative
to the gyroscopes' axes. 
A homogeneous gravitomagnetic field can be transformed away completely 
by going to FIDOs 
which are nonrotating relative to 
the spin axes of local gyroscopes, i.e.
in homogeneous rotation 
relative to the previous FIDOs.
Physics in a homogeneous gravitomagnetic field
is 
    {\it equivalent} 
to physics on a merry-go-round in Minkowski space.---
The centrifugal force is quadratic in $\Omega,$ 
therefore it is missing in linear perturbation theory.


For vorticity perturbations on a 
FRW background 
with $k=0$ in Cartesian comoving LONBs we obtain
\begin{equation}
\frac{1}{a} \frac{d}{dt} (a p_{\hat{i}})=
\varepsilon [\vec{E}_{\rm g}+
(\vec{v} \times  \vec{B}_{\rm g})    
+ H \vec{v} \times (\vec{\beta} \times \vec{v})]_{\hat{i}}.
\end{equation}
For FRW with $k=\{ 0, \pm 1 \}$ in spherical comoving LONBs  
there are additional terms from
$\partial_{\hat{i}}L_{\hat{j}}$ in 
     Eq.~(\ref{omega.ijk}).
These terms are present even in the absence of
vorticity perturbations and of Hubble expansion, 
because in a spherical basis for $k = 0$ spatial LONBs are not
parallelized, 
and for $k = \pm 1$ global parallelization is impossible 
on $\Sigma_t.$


\section{Einstein's Equations for Gravitomagnetism: Amp\`ere's Law}


The computation of curvature in Cartan's formalism: 
We give a short, self-contained
derivation of the formulae needed from Cartan's formalism.
Readers not interested in calculational methods 
can skip the first part of this section
and go directly to the results starting with 
    Eq.~(\ref{Ampere}).


\subsection{Cartan's second equation in LONB components}


The curvature 2-form 
${\tilde{\cal R}}^{\hat{a}}_{\ \hat{b}}$
has LONB components 
$({\cal R}^{\hat{a}}_{\ \hat{b}})_{\hat{c}\hat{d}},$
which are the LONB components of the 
Riemann tensor $R^{\hat{a}}_{\ \hat{b}\hat{c}\hat{d}}.$
The Riemann tensor can be operationally defined 
by the action of 
$(\nabla_{\gamma} \nabla_{\delta}
- \nabla_{\delta} \nabla_{\gamma})$
on the LONB 1-form
$\tilde{\theta}^{\hat{a}},$
\begin{equation}
(\nabla_{\gamma} \nabla_{\delta}
- \nabla_{\delta} \nabla_{\gamma})
\tilde{\theta}^{\hat{a}} =
- \tilde{\theta}^{\hat{b}} 
({\cal R}^{\hat{a}}_{\ \hat{b}})_{\gamma\delta},
\end{equation}
where the covariant derivatives $\nabla_{\gamma}$
and $\nabla_{\delta}$
must be in the coordinate basis.
To compute the curvature 2-form we first use
$\nabla_{\delta} \tilde{\theta}^{\hat{a}}=
-(\omega^{\hat{a}}_{\ \hat{e}})_{\delta} 
\tilde{\theta}^{\hat{e}}.$
Acting on the right-hand side 
by $\nabla_{\gamma}$ gives two terms.
One term comes from $\nabla_{\gamma}$ acting on 
$\tilde{\theta}^{\hat{e}},$ 
and after antisymmetrization in $[\gamma \delta]$
it produces 
$-(\omega^{\hat{a}}_{\ \hat{e}} \wedge 
\omega^{\hat{e}}_{\ \hat{b}})_{\gamma\delta} 
\tilde{\theta}^{\hat{b}}.$
The other term comes from $\nabla_{\gamma}$ acting on
the expansion coefficient (number field) 
$(\omega^{\hat{a}}_{\ \hat{e}})_{\delta},$ 
where it can be replaced by $\partial_{\gamma},$
and after antisymmetrization in $[\gamma \delta]$
it produces 
$-(d \omega^{\hat{a}}_{\ \hat{b}})_{\gamma \delta}
\tilde{\theta}^{\hat{b}}.$
Hence we obtain
\begin{equation}
\tilde{{\cal R}}^{\hat{a}}_{\ \hat{b}} =
\tilde{d} \tilde{\omega}^{\hat{a}}_{\ \hat{b}} +
\tilde{\omega}^{\hat{a}}_{\ \hat{e}} \wedge 
\tilde{\omega}^{\hat{e}}_{\ \hat{b}},
\end{equation}
which is Cartan's second equation.


To obtain the LONB components of 
the first term of the right-hand side,
$(d\omega^{\hat{a}}_{\ \hat{b}})_{\hat{c}\hat{d}},$
we must first convert the connection components of 
Eqs.~
   (\ref{dbeta/dt} - \ref{omega.ijk}) 
from the LONB to the coordinate basis,
$(\omega^{\hat{a}}_{\ \hat{b}})_{\delta}=
 (\omega^{\hat{a}}_{\ \hat{b}})_{\hat{d}}
(\theta^{\hat{d}})_{\delta},$   
then take the exterior derivative,
$\partial_{\gamma}\{(\omega^{\hat{a}}_{\ \hat{b}})_{\hat{d}}
(\theta^{\hat{d}})_{\delta}\} - [\gamma \leftrightarrow \delta],$   
and then convert the result 
back from coordinate components to LONB components.
The partial derivative of the product gives two terms,
one with a partial derivative of 
$(\omega^{\hat{a}}_{\ \hat{b}})_{\hat{d}},$
the other with $\partial_{\gamma} (\theta^{\hat{d}})_{\delta},$
which produces another connection-1-form component.
In the second term of Cartan's second equation these conversions 
from LONB to coordinate basis and back again cancel, 
since there is no derivative in between.
The result is Cartan's 2nd equation in LONB components,
\begin{eqnarray}
({\cal R}^{\hat{a}}_{\ \hat{b}})_{\hat{c}\hat{d}} &=&
[(e_{\hat{c}})^{\gamma}
\partial_{\gamma} (\omega^{\hat{a}}_{\ \hat{b}})_{\hat{d}} 
- (\omega^{\hat{a}}_{\ \hat{b}})_{\hat{f}} 
(\omega^{\hat{f}}_{\ \hat{d}})_{\hat{c}}
\nonumber 
 + (\omega^{\hat{a}}_{\ \hat{e}})_{\hat{c}} 
  (\omega^{\hat{e}}_{\ \hat{b}})_{\hat{d}}] \\
&&  - [\hat{c} \leftrightarrow \hat{d}].
\label{curvature.computation}
\end{eqnarray}


\subsection{Einstein equations for vorticity perturbations 
of Minkowski space}


For linear vorticity perturbations of Minkowski space
(with Cartesian coordinates for 3-space)
all non-zero connection coefficients in
Eqs.~(\ref{dbeta/dt} - \ref{omega.ijk}) are of first order
in the perturbations. 
Therefore the second term of Cartan's second equation
can be neglected, and in the first term 
one need not distinguish 
components in LONB from components in the coordinate basis. 
For vorticity perturbations the important Einstein equation 
is the equation for
$G_{\hat{0}\hat{i}}=R_{\hat{0}\hat{i}},$
\begin{equation} 
R_{\hat{0}\hat{i}}=
({\cal R}_{\hat{0}\hat{j}})_{\hat{i}\hat{j}}=
(d\omega_{\hat{0}\hat{j}})_{\hat{i}\hat{j}}=
\frac{1}{2}({\rm curl} \vec{B})_{\hat{i}}.
\end{equation}
Hence Einstein's $G_{\hat{0}\hat{i}}$-equation 
for general, i.e. time-dependent, vorticity perturbations 
on a Minkowski background reads
\begin{equation}
{\rm curl} \vec{B_{\rm g}} = -16 \pi G_{\rm N} \vec{J}_{\varepsilon}.
\label{Ampere}
\end{equation}
In contrast to the Amp\`ere-Maxwell equation 
of ordinary electrodynamics,
the Maxwell term
$(\partial_{t} \vec{E})$ is 
    {\it absent} 
in the 
    {\it time-dependent}, 
i.e. non-stationary context of gravitomagnetodynamics.
    Eq.~(\ref{Ampere})
is consistent without $(\partial_{t} \vec{E}_{\rm g}),$ 
because in the vorticity sector
all vector fields are divergence-free,
particularly div$ J_{\varepsilon}=0.$
A term $\partial \vec{E}_{\rm g} / \partial t$
cannot be present in gravitomagnetism, 
otherwise
    Eq.~(\ref{Ampere})
together with Faraday's law for gravitomagnetism,
    Eq.~(\ref{Faraday}),
would erroneously predict gravitational vector waves.---
    Eq.~(\ref{Ampere}) 
for 
    {\it gravito-magneto-dynamics}
(i.e. for a non-stationary context) 
is identical to the 
original law of Amp\`ere for stationary magnetism, 
except that the charge current $\vec{J}_{q}$ 
is replaced by the energy current $\vec{J}_{\varepsilon},$ 
and the prefactor $4 \pi$ is 
replaced by the prefactor $(-16 \pi G_{\rm N}).$---
The source $J_{\varepsilon}^{\hat{i}} \equiv T^{\hat{0}\hat{i}}$ 
is the energy current density, which is equal 
to the momentum density (for $c=1$).
In linear perturbation theory for a perfect fluid  
$\vec{J}_{\varepsilon} = (\rho + p) \vec{v}.$


The $G_{\hat{i}\hat{0}}$ equation 
is an equation at fixed time, 
a constraint equation, called {\it momentum constraint}, 
since the momentum density appears on the
right-hand side of 
    Eq.~(\ref{Ampere}).
To see the analogous structures of gravitomagnetism 
and electromagnetism, it is more instructive to formulate
this constraint equation, as we have done in 
    Eq.~(\ref{Ampere}), 
via connection 1-forms, which involves 
$(\partial_{i} \beta_{j} - \partial_{j} \beta_{i}),$
i.e. the gravitomagnetic field,
than via the extrinsic curvature tensor $K_{ij},$
which involves
$(\partial_{i} \beta_{j} + \partial_{j} \beta_{i}).$
Of course the resulting constraint, if written in terms of
$\vec{A}_{\rm g}=\vec{\beta},$ is the same, 
$\Delta \vec{A}_{\rm g}= 16 \pi G_{\rm N} \vec{J}_{\varepsilon}.$


   The $\delta G_{ \hat{0} \hat{0} }$ equation with
the source $\delta T_{ \hat{0} \hat{0} }$ is trivially fulfilled,
since these two objects are 3-scalars and therefore 
vanish in the vector sector.---
The source $\delta T_{ \hat{i} \hat{j} }$ 
vanishes (for linear vorticity perturbations and perfect fluids), 
since it is of second order in the perturbation.
The $\delta G_{ \hat{i} \hat{j} }$ equations give 
$\partial_{0} (\partial_{i} \beta_{j} + \partial_{j} \beta_{i})=0,$
i.e. the shear of the field 
$\vec{\beta}$ has vanishing time-derivative.


\subsection{The momentum constraint for vorticity perturbations 
of spatially flat FRW spaces}


With Cartesian comoving coordinates for flat 3-space
there are two new terms in the connection coefficients,  
$(\omega_{\hat{i}\hat{0}})_{\hat{j}}^{\rm FRW}=
\delta_{\hat{i}\hat{j}}H$ and 
$(\omega_{\hat{i}\hat{j}})_{\hat{k}}
= H( \delta_{\hat{i}\hat{k}}  \beta_{\hat{j}} -
     \delta_{\hat{j}\hat{k}}  \beta_{\hat{i}} ).$
Computing  
$R_{\hat{0}\hat{i}}=
({\cal R}_{\hat{0}\hat{j}})_{\hat{i}\hat{j}}$
with 
   Eq.~(\ref{curvature.computation})
we obtain the corresponding Einstein equation,
\begin{equation}
{\rm curl} \vec{B}_{\rm g} - 4 (dH/dt) \vec{A}_{\rm g} 
= - 16 \pi G_{\rm N} \vec{J}_{\varepsilon},
\label{Ampere.FRW}
\end{equation}
where we have used $\vec{\beta}=\vec{A}_{\rm g}.$ 
The 
    {\it scale factor} 
$a$ of the spatially flat FRW universe
    {\it does not appear} 
in this fixed-time equation.
Note that 'fixed-time equation' means that there are no
partial time-derivatives of the fields 
$\vec{B}_{\rm g}(\vec{x},t),$ etc. The coefficient $(dH/dt)$
is merely a given input number at each time.--- 
Every symbol in this equation refers to the physical scale.
This is a general fact: If we consider any fixed-time equation 
at any time $t_1,$ we are free to set $a(t_1) = 1,$
hence the scale factor disappears.


Friedmann's equations give $H$-dot,
\begin{equation}
dH/dt =-4\pi G_{\rm N} (\rho + p).
\label{Friedmann}
\end{equation}
Since $(dH/dt) \leq 0$ for
$(p/ \rho) \geq -1,$ we define the $H-$dot radius $R_{dH/dt}$ 
by $R_{dH/dt}^2 = (-dH/dt)^{-1}$, and we define $\mu$ by
\begin{equation}
\mu^2  \equiv  -4 (dH/dt) \equiv 
4 (R_{dH/dt})^{-2}.
\label{mu} 
\end{equation}
%
We use $\vec{B}_{\rm g}=\rm{curl} \, \vec{A}_{\rm g},$  
and with $\rm{div} \,\vec{A}_{\rm g} =0$ we have 
$( {\rm curl~curl}~\vec{A}_{\rm g}) = - \Delta \, \vec{A}_{\rm g}.$ 
Therefore Eq.~(\ref{Ampere.FRW}) becomes
\begin{equation}
\left(- \Delta \,
+\,\mu^2\right) \vec{A}_{\rm g} \;=\; - 16\,\pi\; G_{\rm N}
\;\vec{J}_\varepsilon~, 
\label{Ampere.FRW.A}
\end{equation}
an elliptic equation.
Concerning the absence of partial time-derivatives of
$\vec{A}_{\rm g}$ in 
    Eq.~(\ref{Ampere.FRW.A}) 
for the time-dependent context of gravitodynamics, 
see the comments after 
    Eq.~(\ref{Ampere}).---
The new term on the left-hand side, 
$(- 4 \dot{H}\;\vec{A}_{\rm g})=(\mu^2 \vec{A}_{\rm g})$, 
is responsible for the 
     {\it exponential cutoff} 
on scales larger than the $H$-dot radius, 
     Eqs.~(\ref{Mach.summary},
           \ref{Mach.average.introduction}),
which is crucial to remove the problem of 
     {\it `overdragging'},
as shown in Fig.~1,
and for obtaining a
     {\it weight function with normalization to unity}
in our 
     Eq.~(\ref{Mach.average.introduction}),
which gives $\vec{\Omega}_{\rm gyro}$ 
as the weighted average of $\vec{\Omega}_{\rm matter}.$


Bardeen 
    \cite{Bardeen}
gives
    {\it two alternative choices}
for 
    {\it gauge-invariant amplitudes} 
for the 
    {\it matter velocity},
$v_{\rm s}$ and $v_{\rm c}$
in his Eqs.~(3.21, 3.23).--- 
Bardeen uses Einstein's momentum constraint in
    {\it coordinate-basis components,}
$T^0_j$ on the source side, and this is 
directly proportional to $v_{\rm c}.$---
In contrast we use the momentum constraint in
    {\it LONB components,} 
$T_{\hat{0} \hat{j}},$ and this is 
directly proportional to 
    {\it Bardeen's other gauge-invariant variable} 
$v_{\rm s}.$


The physical meaning 
of $v_s$ and $v_c$:
The field with the amplitude $v_{\rm c},$ used by Bardeen, 
is directly related (via its curl)
to the angular velocity field of matter   
    {\it relative to axes of local gyroscopes all over} $\Sigma_t.$ 
But this has the severe drawback that $v_{\rm c}$ is 
    {\it not measurable without prior knowledge} 
of the solution of Einstein's equations,  
$g_{\mu \nu}$ and therefore $\vec{A}_{\rm g}$ and 
$\vec{B}_{\rm g}.$ 
Hence $v_{\rm c},$ employed by Bardeen, 
   {\it cannot be used as a matter input}
for solving Einstein's equations 
in the context of Mach's principle and of dragging, 
where a measured matter input is needed before
having solved Einstein's equations. 
See also sects. VIII and IX.---
In contrast     
     {\it we use Bardeen's other gauge-invariant amplitude} 
$v_{\rm s},$
which is directly related to 
the angular velocity of matter
     {\it relative to the asymptotic unperturbed quasars}.
This is an input which is
     {\it measurable without prior knowledge} of 
$g_{0 i}= A^{(\rm g)}_i,$
i.e. before knowing the output 
of solving Einstein's constraint equation.


The difference
between the two gauge-invariant amplitudes for the matter velocity
is given by $(v_{\rm c} - v_{\rm s}) = - \Psi,$ 
     Eq.~(3.23) of Bardeen.
The $\Psi$-term, a geometric term,
which is on the right-hand side of Bardeen's momentum constraint,
must be 
     {\it moved to the left-hand side}, 
the geometric side.
With the prefactors in the momentum constraint 
and in our notation 
the $\Psi$-term appears as $\mu^2 \vec{A}_{\rm g}.$


Both sides of our 
     Eq.~(\ref{Ampere.FRW.A})
are 
     {\it gauge-invariant}, 
the left-hand side because
our $\vec{A}_{\rm g}$ is directly proportional to 
Bardeen's gauge-invariant potential $\Psi,$
    Eq.~(\ref{beta.equiv.Psi}),
the right-hand side because
our matter velocity $\vec{v}$ is directly proportional to
Bardeen's gauge invariant amplitude $v_{\rm s}.$


\section{Mach's Principle}

\subsection{Evolution of inertial axes determined 
exclusively by cosmic matter flows}


The solution of 
     Eq.~(\ref{Ampere.FRW.A})   
is the Yukawa potential for
$\vec{A}_{\rm g}=\vec{\beta}$
in terms of the energy currents~$\vec{J}_\varepsilon$ 
at the same time,
\begin{equation}
\vec{A}_{\rm g} (\vec{r},t)=
-4\, G_{\rm N} \int d^3
r'\,\vec{J}_\varepsilon ( \vec{r'}, t)~ \frac{{\rm
    exp}( - \mu |\vec{r}-\vec{r'}|)}{|\vec{r}-\vec{r'}|}.
\label{Yukawa.potential}
\end{equation}
This is analogous to the formula for ordinary magnetostatics 
except for the exponential cutoff,
i.e. the $1/r$ potential is replaced 
by   the Yukawa potential \,
$(1/r) {\rm exp} (- \mu r).$
The Green function which is exponentially growing for 
$r' \rightarrow \infty$ has been rejected on the standard grounds 
of field theory.
This gives our first conclusion: 
The contributions of vorticity perturbations
beyond the $H$-dot radius are exponentially suppressed.


The fundamental law of gravitomagnetism in integral form 
for linear vorticity perturbations and 
for spatially flat FRW  universes ($k=0$),
the equation for 
$\vec{B}_{\rm g}$  and  $\vec{\Omega}_{\rm gyro}$
in terms of 
sources at the same time, is
\begin{eqnarray}
\vec{B}_{{\rm g}} (P) &=&  - 4G_{\rm N} \int d^3 r_Q \, 
[\vec{n}_{PQ} \times \vec{J}_{\varepsilon}(Q)]  \, 
Y_{\mu} (r_{PQ})
\label{B.general}
\\
Y_{\mu} (r) &=& \frac{-d}{dr} \,  \frac{e^{- \mu r}}{r}
= \mbox{Yukawa force}, 
\label{Yukawa.force}
\\
&& \quad \vec{\Omega}_{\rm gyro} 
\,=\, - \frac{1}{2} \vec{B}_{\rm g}(\vec{r}_{\rm gyro}), 
\label{Omega.B.last.eq}
\end{eqnarray}   
where $\vec{n}_{PQ}$ is the unit vector pointing 
along the geodesic on $\Sigma_t$ 
from $P$ to $Q,$
and $Q$ stands for ``source point''. 
We have applied  
$\vec{B}_{\rm g}={\rm curl} \, \vec{A}_{\rm g}$ to 
   Eq.~(\ref{Yukawa.potential}).


Both $\vec{B}_{\rm g} = - 2 \vec{\Omega}_{\rm gyro}$ 
and the transverse velocity in
$[\vec{n}_{PQ} \times \vec{J}_{\varepsilon}(Q)]$ are 
{\it measured} relative to
{\it geodesics} on $\Sigma_t$ from the the gyroscope at $P$ 
to {\it quasars} in the asymptotic FRW universe.
Since $\Sigma_t$ is a Euclidean 3-space, 
global parallelism on $\Sigma_t$ can be used
when comparing vectors at $P$ and $Q$.
Although we use Cartesian coordinates for Euclidean 3-space,
all quantities in 
   Eq.~(\ref{B.general})
have a {\it gauge invariant} meaning,
they are directly measurable,
e.g. $d^{\, 3}r_Q$ is the measured volume element,
and $r_{PQ}$ is the geodesic distance from $P$ to $Q.$---
The scale factor $a$ of the spatially flat FRW universe
does not appear in the 
    fixed-time equation (\ref{B.general}),
as discussed after 
    Eq.~(\ref{Ampere.FRW}); 
every symbol refers to the physical scale.


The only differences from 
Amp\`ere's law in integral form are:
\begin{enumerate}
\item the replacement of the current of electric charge $\vec{J}_q$
      by the measured energy current of matter
      $\vec{J}_{\varepsilon},$
\item the factor $(- G_{\rm N}),$ as in the transition from 
      Coulomb's law to Newton's law, 
\item the additional factor $4,$
      which occurs in the transition from Amp\`ere's law of 
      ordinary magnetism to gravitomagnetism, 
\item the replacement of the $1/r^2$ force in Amp\`ere's law
      by the Yukawa force with its exponential cutoff,
      Eq.~(\ref{Yukawa.force}),
      which occurs in the transition from a Minkowski
      background to gravitomagnetism on a a 
      FRW background with $k=0.$ 
\end{enumerate}
%


An analogous Yukawa cutoff at super-$H$-dot scales
in the 
     {\it scalar sector} 
of cosmological perturbation theory occurs in the
'uniform expansion gauge', 
which is also called the 'uniform Hubble gauge';
see 
    Ref.~\cite{Schmid.Schwarz.Widerin}.


Note the 
   {\it fundamental difference} 
between 
cosmological gravitomagnetism,
    Eq.~(\ref{B.general}),
and Amp\`ere's law.
The latter does not hold in a rotating reference frame,
unless one introduces fictitious forces.
The same is true for the equations for electromagnetism
in special relativity,
and for general relativity of the solar system in asymptotic
Minkowski space, where fictitious forces must be encoded
by boundary conditions at spatial infinity
as explained in Sec.~XI.
In contrast, for cosmological gravitomagnetism,
    Eq.~(\ref{B.general})
remains valid, as it stands, in a frame which is rotating
with angular velocity $\vec{\Omega}^{*}$ relative to asymptotic quasars,
as explained in sec. II after
    Eq.~(\ref{Mach.summary}).
This establishes the fact that 
the asymptotic inertial frame has no influence
in cosmological gravitomagnetism.
The local nonrotating frame 
(at any point $P$)
is not needed as an input in applying
    Eq.~(\ref{B.general}),
i.e.      
    {\it no absolute element} 
is needed as an input.
The time evolution of local inertial axes is an output, 
    {\it determined exclusively}
by the weighted average of cosmological energy flows.


    Eqs.~(\ref{B.general}) -
         (\ref{Omega.B.last.eq})
state what 
    {\it specific average} 
of the measured energy flow of matter 
out there in the universe determines 
the motion of gyroscope axes here.
This answers Mach's question
    \cite{Mach.Energy.what.share}:
\begin{quotation}
``What share has every mass in the determination  
of direction ... in the law of inertia? 
No definite answer can be given by our experiences.''
\end{quotation}  
%


The gravitomagnetic moment density 
$\vec{\mu}_{\rm g}$ 
     {\it per volume element}
is the analog of the magnetic moment density
(with $\vec{J}_q$ replaced 
by the measured energy current $\vec{J}_{\varepsilon}$).
The gravitomagnetic moment density appears as the source in
    Eq.~(\ref{B.general}), 
and it is equal to one half of the 
measured angular momentum density of matter
$\vec{L} = 
(\rho + p) \, (\vec{r} \times \vec{v})$
at the source point $Q$ relative to the gyroscope point $P,$
\begin{eqnarray}
  \vec{\mu}_{\rm g}   
&=&  \frac{1}{2}  [ \vec{r}_{PQ} \times \vec{J}_{\varepsilon}(Q)]
=  \frac{1}{2} \vec{L}_{PQ},
\label{magnetic.moment.angular.momentum}
\\
\vec{\Omega}_{\rm gyro} (P)
&=& -\frac{1}{2} \vec{B}_{\rm g}(P)
\nonumber
\\ 
&=& 2G_{\rm N} \int d^3 r_Q  \, \,
\frac{\vec{L}_{PQ}}{r_{PQ}}  \, \, 
Y_{\mu} (r_{PQ}).
\label{B.general.L}
\end{eqnarray}
It is a matter of taste and choice of emphasis, 
whether the term 
    `{\it source}' 
for the gravitomagnetic field is used for 
(1) the measured energy current ($=$ momentum density) 
$\vec{J}_{\varepsilon}$
as in the analogue of Amp\`ere's law (momentum constraint),
    Eqs.~(\ref{Ampere}) and
         (\ref{Ampere.FRW}), or
(2) the measured angular momentum density 
$\vec{L} = (\vec{r} \times \vec{J}_{\varepsilon}),$ 
as in the integrated form of the analogue of Amp\`ere's law,
    Eq.~(\ref{B.general.L}).


\subsection{Evolution of inertial axes exactly follows 
the average of cosmic matter flows}


The gravitomagnetic moment density 
   {\it per} $r$-{\it interval},  
$d\vec{\mu}_{\rm g}/dr,$ 
is the lowest multipole term, the $\ell=1 $ term, 
of the ``odd parity sequence'', 
$P = (-1)^{\ell +1},$
in the multipole expansion of 
the vector source $\vec{J}_{\varepsilon}$
for $r_{\rm obs} = r_{\rm gyro}=0$ and $
r_{\rm source} > 0.$ 
Other multipoles cannot contribute 
to the gyroscope's precession, i.e. 
to the gravitomagnetic field $\vec{B}_{\rm g}$
at $ r = 0,$ for reasons of symmetry under rotations 
and space reflection.---
In first-order perturbation theory $(\rho + p) =$constant
all over space, hence also the 
    {\it velocity field} 
at a given radius $r$ can only contribute via the term with  
$\{ \ell = 1, \, \mbox{odd parity sequence} \},$
which is 
    {\it equivalent} 
to a 
    {\it rigid rotation}  
with the angular velocity $\vec{\Omega}_{\rm matter}(r)$.
Using this fact and 
   Eqs.~(\ref{Friedmann},
         \ref{mu})
we obtain
\begin{eqnarray}
\vec{\Omega}_{\rm gyro} &=& 
\frac{\mu^2}{3} \int_0^\infty  dr \, r  \, 
\vec{\Omega}_{\rm matter}(r) \, C_{\mu}(r),
\label{Mach.2}
\\
C_{\mu}(r) &=& r^2 Y_{\mu} (r) = (1 + \mu r) \exp(- \mu r).
\label{cutoff.function}
\end{eqnarray}
This is our most important result.
     Eq.~(\ref{Mach.2})
shows that $\vec{\Omega}_{\rm gyro}$ 
is the weighted average  
of $\vec{\Omega}_{\rm matter},$  
i.e. the precession of 
a gyroscope, $\vec{\Omega}_{\rm gyro}$,
         {\it exactly follows,}  
i.e. it is 
         {\it exactly dragged}
by the weighted average 
of $\vec{\Omega}_{\rm matter}.$---
The weight function is 
     {\it normalized to unity}, 
as it must be for an averaging weight function in any problem.
This crucially depends on the 
     {\it exponential cutoff}
in the Yukawa force $Y_{\mu}(r).$
The cutoff function 
$C_{\mu}(r)$ is $1$ for $r \ll \mu^{-1}$ 
and goes to zero exponentially fast for $r \gg \mu^{-1}.$
The weight function 
per logarithmic r-interval in 
   Eq.~(\ref{Mach.2})
is $r^2 C_{\mu}(r),$ i.e. it 
grows quadratically
until one reaches, roughly, the $R$-dot radius,
and then it goes to zero exponentially.---
The conclusion is that    
   Eq.~(\ref{Mach.2})
is a clear demonstration of how 
 {\it Mach's principle is contained within cosmological general relativity}.


When talking about the ``angular velocity of cosmic matter''
one must keep the following in mind:
For a given arbitrary velocity field 
the densities of gravitomagnetic moment 
and of measured angular momentum 
at  
   {\it one source point} $Q$ 
relative to the gyroscope point $P$
are well-defined. 
But one cannot uniquely define 
a density of angular velocity
at 
   {\it one} 
source point $Q$ 
relative to $P, \, \, \vec{\Omega}(Q;P),$ 
because one can add terms $\vec{r}_{PQ} f(Q)$
without changing $\vec{v}(Q).$


\subsection{The dragging fraction}


The dragging fraction
refers to the simple cosmological model
discussed in the Introduction,
a uniform (rigid) rotational motion
with $\vec{\Omega}_{\rm matter}$ 
constant inside a radius
$R_{\rm rot}$ 
around a gyroscope and zero outside
and with
energy density and pressure  
spatially constant everywhere.
The definition of the dragging fraction $f_{\rm drag}$   
is given by
\begin{equation}
\vec{\Omega}_{\rm gyro} \, 
= \, f_{\rm drag} \, \, \vec{\Omega}_{\rm matter},
\end{equation}
where  $\vec{\Omega}_{\rm gyro}$ and 
$\vec{\Omega}_{\rm matter}$
are measured relative to geodesics on $\Sigma_t$ 
from the gyroscope
to quasars in the asymptotic FRW space.---
A better word than dragging fraction is 
voting power of the rotating matter 
(in contrast to the voting power 
of the non-rotating matter outside), 
because there is exact dragging of 
$\vec{\Omega}_{\rm gyro}$ by a weighted average of
$\vec{\Omega}_{\rm matter}$ according to 
    Eq.~(\ref{Mach.2}).---
Putting  $\vec{\Omega}_{\rm matter}=$ constant 
inside a radius $R_{\rm rot}$ 
and  $\vec{\Omega}_{\rm matter} = 0$ outside
in
    Eq.~(\ref{Mach.2})
we obtain the dragging fraction,
\begin{eqnarray}
&& f_{\rm drag} (R_{\rm rot}) 
= 
\frac{\mu^{2}}{3}  \int_0^{R_{\rm rot}} 
dr \, r \, (1 + \mu r) \exp (-\mu r)
\nonumber
\\
&& =
1 - \{ \exp (- \mu r)\, [1 + \mu r + \mu^2 r^2/3] \}_{r=R_{\rm rot}}. 
\label{dragging.result}
\end{eqnarray}
The dragging fraction (voting power of the matter inside $R_{\rm rot}$)
is shown by the solid curve in 
    Fig.~(1).
It grows quadratically with $R_{\rm rot}$ 
until it reaches order of one for $R_{\rm rot}= R_{dH/dt}.$
As $R_{\rm rot}$ grows beyond $R_{dH/dt},$
the dragging fraction approaches the exact value $1$ exponentially fast,
i.e. exact dragging of the gyroscope axes by the rotating matter.
This holds for any equation of state.


The problem of 
     {it `overdragging',}
shown by the dashed curve in
     Fig.~(1), 
arises when extrapolating the results of 
gravitomagnetic perturbation theory on a Minkowski background
beyond the region of validity, as noted in 
    \cite{MTW.Mach}.---
The solid line shows, how the problem of 
`overdragging' is 
removed by the 
exponential cutoff 
from cosmological perturbation theory for
super-Hubble-dot scales, i.e. 
by the exponential cutoff in
    Eqs.~(\ref{Ampere.FRW.A},
          \ref{B.general}).---
Lynden-Bell et al 
    \cite{Lynden-Bell.1995}
do not have an exponential cutoff,
because they use the 
conserved 
angular momentum instead of the 
    {\it measured} 
angular momentum as the source. 
See our discussion below in 
    sect.~\ref{Measured.Matter.Input}.


   {\it Analogous anti-dragging effects in magnetostatics:}
A rotating charged spherical shell acting on  
magnetic dipole moments inside 
has the opposite sign in Amp\`ere's law compared to
gravitomagnetism, 
    Eq.~(\ref{Ampere}),  
and this causes antidragging in the magnetostatic case.


{\it Physics could have evolved differently after Mach} 
(1883):
Physicists at that time could have extended the correspondence
between Coulomb's law and Newton's law and postulated
gravitomagnetism in correspondence to Ampere's magnetostatics.
Their prediction of the dragging of inertial frames would
have been too small by a factor $4$ compared to the correct result 
for linear gravitomagnetism on a Minkowski background.
For the simple case of a homogeneous rotation out to some radius $R$
of cold matter with $\rho =$ constant
they would have 
obtained perfect dragging for
$GM/(Rc^2) = 2$ and 
noted the problem of `overdragging' 
for $GM/(Rc^2) > 2.$ 
In the late 1920's, for an approximately 
homogeneous and expanding universe
with cold matter,
this critical radius would have been identified with the Hubble radius
(apart from a missing factor $2).$


Measuring everything relative to axes of gyroscopes 
at one given location 
(instead of relative to quasars in the asymptotic FRW space)
makes the the left-hand side of 
   Eq.~(\ref{B.general.L}) 
vanish. Hence 
   Eq.~(\ref{B.general.L}) 
reduces to the statement that the 
measured angular momentum 
of matter 
relative to gyroscopes at one given location 
will vanish after averaging  
with the weight $r^{-3}$ and with the 
exponential cutoff $C_{\mu}(r).$---
From the point of view of measurements
it is preferrable to mesure relative to gyroscopes at one given location.
But to see the structure of gravitomagnetism 
and its correspondence with electromagnetism most clearly,
it is best to measure relative to quasars in the asymptotic FRW space.


\subsection{The de Sitter limit}


Dark energy with $p/ \rho = - 1,$    
i.e. a cosmological constant,
does not contribute in Mach's principle,
   Eq.~(\ref{B.general}),
since there is no flow of energy 
associated with it, its energy current 
$\vec{J}_{\varepsilon}=(\rho + p) \vec{v}$ vanishes.---
When considering a FRW universe with $k=0$ 
and letting $(p / \rho)$  get closer and closer to $(-1)$ 
at fixed Hubble rate (hence fixed $\rho$) 
until we are 
  {\it arbitrarily close to a de Sitter universe,}
  {\it Mach's principle continues to work} 
(in linear perturbation theory).
When taking this limit the prefactor 
$\mu^2 = (16 \pi G_{\rm N}) (\rho +p)$ in 
    Eq.~(\ref{dragging.result})
gets smaller and smaller (and tends to zero), 
but the $H-$dot radius $\mu^{-1},$
which cuts off the otherwise quadratically divergent integral in
    Eq.~(\ref{dragging.result}),
gets larger and larger such that perfect dragging is maintained
for arbitrarily small $\mu^2,$ 
i.e. arbitrarily close to a de Sitter universe.


\subsection{Einstein's objection: Interaction of masses}


Einstein's objection to Mach's principle in his 
autobiographical notes of 1949 
     \cite{Einstein}:
\begin{quotation} 
"Mach conjectures that inertia would have to depend
upon the interaction of masses, precisely as was true for Newton's
other forces, a conception which for a long time I considered as in
principle the correct one. It presupposes implicitly, however, that
the basic theory should be of the general type of Newton's mechanics:
masses and their interactions as the original concepts.  The attempt
at such a solution does not fit into a consistent field theory, as
will be immediately recognized."
\end{quotation}
We have shown how this apparent difficulty
is resolved in General Relativity, 
specifically in linear Gravitomagnetism:
The relevant Einstein equation (the momentum constraint) 
has the form of Amp\`ere's law with a Yukawa cutoff, 
   Eq.~(\ref{Ampere.FRW}). 
It is a remarkable fact that Einstein's 
   {\it local field equation}
for $G_{\hat{0} \hat{i}}$ 
has the solution
   Eq.~(\ref{B.general}), 
which has the form of an 
   {\it instantaneous action-at-a-distance}.
It says that the measured mass-energy flow out there in the universe 
does indeed determine the precession of gyroscope axes here.--- 
See also refs.
   \cite{Lindblom.Brill}
   \cite{instantaneous.inertial.frame}.


\section{Measured Matter Input for Einstein's Equations}
\label{Measured.Matter.Input}


The     
        {\it input} 
on the right-hand side of 
   Eqs.~(\ref{B.general},
         \ref{Mach.2}),
which express Mach's principle, is the 
measured angular velocity of matter (stars, galaxies, etc) 
relative to the geodesics on $\Sigma_t$ 
to quasars in the asymptotic FRW space.
        {\it No knowledge of the metric perturbation}
$\vec{\beta}$ is 
        {\it needed} when determining the input for 
   Eqs.~(\ref{B.general},
         \ref{Mach.2}).
The measured angular velocity of matter is a purely 
kinematic or kinetic input.
Similarly the measured angular momentum density 
$\vec{L} = 
(\rho + p) \, (\vec{r} \times \vec{v}) \, = \,
 (\rho + p) \, ( r^{2} \, \sin^2 \theta \, \, \vec{\Omega})$
is a purely kinetic input. We call it the
        {\it kinetic angular momentum}.--- 
The     
        {\it geometric-dynamical output} 
of solving Einstein's $G_{\hat{0}\hat{i}}$ equation is  
$\vec{\beta}= \vec{A}_{\rm g},$ 
the gravitomagnetic vector potential in
    Eqs.~(\ref{Ampere.FRW.A},
          \ref{Yukawa.potential}), 
hence the 
gravitomagnetic force 
$\vec{B}_{\rm g}$ in
     Eq.~(\ref{B.general}).---
The measured, kinetic angular momentum must be 
distinguished from the 
         {\it canonical angular momentum}, 
which we introduce 
in the following paragraphs.


\subsection*{Kinetic versus canonical momentum and angular momentum}


The 
     {\it Einstein-Hilbert action} 
for linear vorticity perturbations (i.e. gravitomagnetism)
on a Minkowski background and for point particles gives
\begin{eqnarray}
S  &=&  (16 \pi G_{\rm N})^{-1} \int d^{4}x \, \sqrt{g} \, \, 
           ({\rm curl} \, \vec{A}_{\rm g})^{2} \nonumber \\
   &+& \int dt \sum_{n} \, [\frac{1}{2} m \, \dot{\vec{x}}_{n}^{2} 
        + m \, \dot{\vec{x}}_{n} \, \vec{A}_{\rm g}(\vec{x}_{n},t)]. \quad
\label{action}
\end{eqnarray}
It is natural to take the nonrelativistic expression ($v \ll c$)
for matter when discussing linear perturbations.
Except for the prefactor $-(16 \pi G_{\rm N})^{-1}$ in the first term 
and the prefactor $m$ in the last term, 
the action is the same as for electromagnetism without
the $(\partial_{t} \vec{A} \, )^{2}$ term.

The 
      {\it Lagrangian} 
gives the equations of motion via the 
      {\it standard} 
Euler-Lagrange equations 
(as in classical mechanics and classical electrodynamics),
if and only if the Lagrangian is defined by 
\begin{equation}
S = \int dt L
\end{equation}
without any metric factors in the integrand.
Hence the Lagrangian for a point particle 
in a gravitomagnetic field is given by the
square bracket of the matter term in 
      Eq.~(\ref{action}).


The     
      {\it canonical momentum} 
of Lagrangian mechanics 
is defined by 
$(p_{\rm can})_{k} = \partial L / \partial \dot{x}^{k},$ 
where $k = 1, 2, 3.$ From 
    Eq.~(\ref{action}) 
we obtain in Cartesian coordinates 
\begin{equation}
\vec{p}_{\rm can} = m (\dot{\vec{x}} + \vec{A}_{\rm g}).
\label{can.mom}
\end{equation}
This is the same equation as in classical Lagrangian mechanics 
for point particles in an electromagnetic field,
except that the electric charge $q$ is replaced by the mass $m$.---
The       
          {\it kinetic momentum} 
is $m\dot{\vec{x}}.$
It can be used as the 
          {\it input} 
for solving Einstein's equations, 
because it is directly determined by measurements
          {\it without prior knowledge}
of the gravitomagnetic vector potential 
$\vec{A}_{\rm g} =  \vec{\beta},$ which is an 
          {\it output}
of solving Einstein's equations. 
On the other hand the 
canonical momentum 
(for a given measured state of motion) depends on the 
gravitomagnetic vector potential $\vec{A}_{\rm g}.$
Therefore the canonical momentum 
          {\it cannot} 
be used as an 
          {\it input} 
for solving Einstein's equations.
The canonical momentum 
          {\it cannot} 
be determined by a FIDO 
from matter measurements without prior knowledge of the 4-geometry.---
In curvilinear coordinates of Euclidean 3-space 
the general definition of the canonical momentum, 
$(p_{\rm can})_{k} = \partial L / \partial \dot{x}^{k},$ gives
\begin{equation}
(p_{\rm can})_{k} = 
m \, \, g^{(3)}_{k n} \, (\dot{x}^{n} + A_{\rm g}^{n}) 
\quad \quad k,n = 1, 2, 3.
\label{can.mom.curvilin}
\end{equation}
From its definition via the Lagrangian,   
the canonical momentum is a 1-form in 3-space, 
i.e. it has a lower 3-index.---  
On the other hand we can also start from the 4-velocity $u^{\nu},$ 
which is the archetype of a 4-vector (tangent vector),  
multiply it with the mass to obtain the 4-momentum $p^{\nu},$ 
and pull down the 4-index
with $g^{(4)}_{\kappa \nu }.$ 
For the spatial components $p_k$ this gives 
$p_{k} = 
m \, \, g^{(4)}_{k \nu} u^{\nu}= 
m \, \, g^{(4)}_{k n} (\dot{x}^{n} + A_{\rm g}^{n})$ 
for $v \ll c,$
which is the same as
the canonical momentum in Eq.~(\ref{can.mom.curvilin}).---
We note that in spherical coordinates
$p_{\phi}$ has the physical meaning of
canonical angular momentum around the $z$-axis.


Mathematically one can freely raise and lower indices 
of the vectors $\bar{p}$ resp. $\bar{v}$
using the metric tensor, 
or convert e.g. from lower indices 
to LONB components
using $p_{\hat{a}} = (e_{\hat{a}})^{\mu} p_{\mu}.$
But physical quantities 
are prototypes for 
either LONB components,               
or     1-form components,
or     contravariant vector components.
For exemple in spherical coordinates the 
     {\it angular velocity} 
around the $z-$axis 
is given by an upper-index component, 
$v^{\phi} = dx^{\phi}/dt,$ the 
     {\it canonical angular momentum} 
around the $z-$axis 
is given by a lower-index component, 
$p_{\phi} = m g_{\phi \mu} dx^{\mu}/dt
= m r^2 \sin^2  \theta \, [(dx^{\phi}/dt) \, + A^{\phi}],$ 
while the 
     {\it measured linear momentum} 
in the direction 
$\vec{e}_{\hat{\phi}}$ 
is given by a LONB component,
$p_{\hat{\phi}} = m \sqrt{g_{\phi \phi}} dx^{\phi}/dt
= m r \sin \theta \, (dx^{\phi}/dt).$ 
The physics question dictates, which type of index 
is the relevant one.


In the special case of azimuthal symmetry, 
i.e. when 
$\bar{e}_{\phi}=\bar{\xi}$ 
is a rotational Killing vector, 
the canonical angular momentum 
$ p_{\phi}=\langle \tilde{p}, \bar{\xi} \rangle $
is conserved, 
while the measured, kinetic angular momentum of matter,
$(r  \, \sin \theta \, p_{\hat{\phi}})$,
is not conserved because of the gravitoelectric induction field
$\vec{E}_{\rm g},$ 
     Eq.~(\ref{Faraday}).
The canonical angular momentum is
relevant for the {\it time-evolution}, i.e. for the {\it dynamics}
and {\it conservation laws}, 
not for kinematics (measurements at a given time).


In the continuum description of matter 
(energy current density)
the LONB components $T_{\hat{0} \hat{k}}$ can be used as 
      {\it input}
for solving Einstein's equations, since they can be measured without
knowing the 
      {\it output} $\beta_{\hat{k}}$ of Einstein's equations.
On the other hand the coordinate-basis components
$T^{0}_{\ k}$ cannot be used as an input, because they cannot be
determined by matter measurements without prior knowledge of $\beta_{k}.$


     In {\it Mach's principle} the input is the 
observed 
     {\it angular velocities} 
of cosmic matter around us as in 
     Eq.~(\ref{Mach.average.introduction})
resp. the measured 
     {\it kinetic} angular momenta
$(\rho + p)(\vec{r} \times \vec{v})$ as in 
    Eq.~(\ref{B.general}).
This has also been recognized by Lynden-Bell et al
    \cite{Lynden-Bell.1995},
who wrote (in section 2.4) ``In Mach's principle, 
we wish to relate the dragging 
not merely 
to the angular momentum distribution but also 
to the distribution of angular velocity 
of the 'stars' about us.''
Unfortunately Lynden-Bell et al
    \cite{Lynden-Bell.1995} 
did not follow up on this important observation, 
i.e. they did not solve the corresponding equation. 
Instead, in the rest of their paper, they used the 
     {\it canonical} 
angular momentum $T^0_{\, \phi}$ as the input
on the right-hand side of Einstein's equation.
For the reasons explained in this section, 
measured kinetic matter 
     {\it input} 
versus dynamical-geometric 
     {\it output} 
of solving Einstein's equations,
we think that their approach would be 
a very large detour
to obtain our 
   Eq.~(\ref{Mach.average.introduction}),
which is the most direct expression of
the dragging of inertial frames.---
If one follows their proposal, 
the $(\dot{H}\vec{A}_{\rm g})-$term on the 
left-hand side of the 
     Einstein equation (\ref{Ampere.FRW}) 
is absent, and there is no exponential suppression factor
multiplying the canonical angular momentum distribution.
This version of the momentum constraint equation 
may be may be relevant in other contexts.---
In response to our conference paper 
    \cite{my.gr-qc.02}
Bi\v{c}\'ak, Lynden-Bell, and Katz
    \cite{Bicak} 
gave arguments why they thought that 
the canonical angular momentum density 
should be considered more fundamental,
although they agreed that 
``the problem with given angular velocities
may be closer to Mach's original principle.''
They solved the equations
for both input possibilities 
(canonical versus kinetic angular momentum) 
for $k = 0, \pm 1.$


\section{Einstein's objection: ``If you have a tensor 
 ${\bf T_{\mu \nu}}$ and not a metric ...''} 
\label{Einstein:if.you.have.tensor}


Einstein's objection to Mach's Principle 
in his letter to Felix Pirani of 2 February 1954  
     \cite{Einstein.Pirani}, 
as quoted by Ehlers in
     \cite{Barbour.Pfister.page93}:
\begin{quotation} ``If you have a tensor $T_{\mu\nu}$ and not a metric,
then this does not meaningfully describe matter.
There is no theory of physics so far, which can describe
matter without already the metric as an ingredient of the 
description of matter. Therefore within existing theories
the statement that the matter by itself determines the metric
is neither wrong nor false, but it is meaningless.''
\end{quotation}
From this argument Einstein drew the conclusion 
that ``one should no longer speak of Mach's principle at all'',
quoted by Renn in
     \cite{Barbour.Pfister.page93}.


We agree with 
Einstein's statement as quoted by Ehlers in
         \cite{Barbour.Pfister.page93}, 
as long as one refers to 
        {\it coordinate-basis components} 
$T_{\mu \nu}.$
Einstein's valid criticism applies to 
$T^{0}_{\ \phi}$ in the proposal of  Bi\v{c}\'ak  et al
        \cite{Bicak}.
On the other hand, for 
        {\it LONB components} 
$T_{\hat{a} \hat{b}}$ 
our conclusions are entirely different.
We have already explained this issue 
in the context of linear gravitomagnetism in the previous section.
We shall now discuss Einstein's objection in a general context.


The metric in LONB components, 
the Lorentz metric
$\eta_{\hat{a} \hat{b}} = \mbox{diag} ( -1, +1, +1, +1 ),$
by itself does
     {\it not encode specific information}           
about the 
     {\it particular}                                         
Riemann space at hand                 
(e.g. perturbed FRW metric versus Schwarzschild metric), 
while the metric in coordinate-basis components 
$g_{\alpha \beta}(x)$ 
encodes all specific information                  
about the particular Riemannian space at hand                 
(curvature etc),                                                 
i.e. it is metric data specific to the particular geometry.    
Both objects, the 
     {\it universal} Lorentz metric $\eta_{\hat{a} \hat{b}}$    
and the 
     {\it case-specific data-set}                             
$g_{\alpha \beta}(x),$  are metrics, 
but they play totally different roles in the context of 
input versus output when solving Einstein's equations.


In Cartan's LONB formalism the 
case-specific data set
which encodes
the specific metric information 
(about the particular Riemann space at hand) 
is obtained by measuring the LONB components 
of the coordinate basis vectors
$\bar{e}_{\mu}$ 
all over the particular Riemann space, 
i.e. measuring $(e_{\mu})^{\hat{a}}_x,$ 
which gives 
$g_{\mu \nu}(x)$ directly via 
$g_{\mu \nu}(x) \, = \,
(e_{\mu})^{\hat{a}}_x \, \, (e_{\nu})^{\hat{b}}_x \, \, 
\eta_{\hat{a} \hat{b}}.$

    
In the LONB formalism a choice must be made 
for a specific field of LONBs,    
$\bar{e}_{\hat{a}}(P).$   
For a coordinate system $P \rightarrow x^{\alpha}_P$ 
given on a manifold 
one class of choices for LONBs is obtained by
an orthonormalization procedure starting from the coordinate basis,
as we have done in
     Eq.~(\ref{expansion of basis vectors}).


Having the universal metric $\eta_{\hat{a} \hat{b}}$ available
is analogous to a surveyor 
having meter sticks available as a universal tool 
before starting the work of surveying
a specific coordinatized two-dimensional Riemann surface. The 
case-specific output data-set 
of the surveyor's work is a table of measurements
of $(e_{\mu})^{\hat{a}}_P,$ hence $g_{\mu \nu}(P)$ 
for a grid of points.


The basic issue is 
     {\it not} 
the fact that the universal metric tool 
$\eta_{\hat{a} \hat{b}}$ makes it 
     {\it possible} 
to perform the work of 
measuring the case-specific data-set 
$g_{\mu \nu}(P)$ all over space-time.
The basic issue is that such measurements are 
     {\it not needed} 
as a case-specific input data-set, 
if one wants to solve Einstein's equations and thereby 
     {\it predict} 
$g_{\mu \nu}(x)$ as a 
case-specific output data-set.


The crucial question 
is: What are the needed case-specific                        
``ingredients'',                                   
what is the needed 
case-specific input data-set                      
for solving Einstein's equations and thereby 
obtaining the case-specific metric output data-set?


The input data set
for solving Einstein's equations is obtained by 
performing the work of measuring the LONB components
of the 4-momenta $\bar{p}$ of all particles in tangent spaces 
all over space-time, i.e. measuring $p^{\hat{a}}.$
Summing over all particles within one measured ${\rm cm}^3$ 
around a point gives
the LONB components $T_{\hat{a} \hat{b}}(P),$
the observed matter input data 
for solving Einstein's equations.
To obtain this matter input 
one only needs 
the universal Lorentz metric 
$\eta_{\hat{a} \hat{b}} = \mbox{diag} \{ -1, +1, +1, +1 \},$
but one does not need 
the case-specific geometric data 
$g_{\alpha \beta}(x),$ 
which are the 
case-specific geometric output data 
from solving Einstein's equations 
and not available at the input level.


We conclude that 
    {\it Einstein's objection} 
quoted above
    {\it  does not apply} 
to 
    {\it LONB components} 
$T_{\hat{a} \hat{b}},$ 
because no case-specific metric data-set 
is needed as an ingredient 
to obtain $T_{\hat{a} \hat{b}}.$---
If one wants to implement the phrase 
``measured matter properties (by themselves) act on space, 
telling it how to curve ''
      \cite{MTW},
one is 
     {\it forced to use LONB components.} 
And for LONBs the most convenient formalim is the one of \'{E}.~Cartan.


\section{Local vorticity measured by non-rotating observers}
\label{local.vorticity}


Local vorticity is defined as ${\rm curl} \, \vec{v}_{\rm fluid}$ 
measured in the local inertial coordinate system which is
comoving with the fluid, i.e. measured relative to the axes of
local gyrosopes (``local compass of inertia''). 
In general coordinates the local vorticity measured 
by non-rotating observers is
\begin{equation}
\omega^{\alpha}= - \varepsilon^{\alpha\beta\gamma\delta}
u_{\beta} \nabla_{\gamma} u_{\delta},
\end{equation} 
where $\varepsilon^{\hat{0}\hat{1}\hat{2}\hat{3}} \equiv -1.$


Mach's principle, formulated as a general hypothesis by Mach
   \cite{Mach.Mechanics.quote}
   \cite{Mach.Energy.what.share} 
and made precise in 
   Eq.~(\ref{B.general}), 
states that the gyroscope axes here follow the rotational energy currents 
$\vec{J}_{\varepsilon}$ of 
matter in the universe 
      {\it averaged} 
with a $r^{-2}$ weight and 
an exponential cutoff at the $H$-dot radius.
In general the gyroscope axes here most definitely do not follow the
motion of the {\it local fluid here}, i.e. relative to gyroscopes
the {\it local vorticity is nonzero according to Mach's principle} 
in general.


There is a special model, a rigid rotation of a fluid 
out to a rotation radius $R_{\rm rot}.$
In the limit $R_{\rm rot}/R_{dH/dt} \rightarrow \infty$
this special model produces an unperturbed FRW universe,
where the local vorticity measured by non-rotating observers vanishes.


Unfortunately many authors have considered the vanishing of the
vorticity relative to the local compass of inertia to be
a test for Mach's principle.
See e.g. Ozsv\`ath and Sch\"ucking's 
     \cite{Ozsvath}
solution and discussion of a Bianchi IX model 
for perturbations of the closed 
and static Einstein universe.
In contrast we conclude that the 
{\it vanishing of the vorticity 
relative to the local compass of inertia} is 
{\it not relevant} as a test of Mach's principle.


The claim of Ozsv\`ath and Sch\"ucking 
      \cite{Ozsvath}
that their Bianchi IX model violates Mach's principle
is incorrect, 
because their claim is based on the non-vanishing of the 
      {\it local vorticity}. 
Now the question is:
Does the Bianchi IX model
of Ozsv\`ath and Sch\"ucking satisfy Mach's principle
(which involves an integral over energy flows in the universe)?


\section{Boundary Conditions for Einstein's Equations 
in Asymptotic Minkowski Space}


For problems in asymptotic Minkowski space
Einstein's equations by themselves 
are not sufficient to determine a solution,
they must be supplemented by boundary conditions 
at spatial infinity.


An example is the Schwarzschild solution.
Consider Earth in asymptotic Minkowski space.
In General Relativity we are free to choose any coordinate system.
We choose spherical coordinates with 
     {\it Earth-fixed axes},
but all the same we write down the standard form of 
the Schwarzschild solution.
This certainly satisfies Einstein's differential equations
for the Earth as a source.
But it is not the correct solution. 
Just consider a geostationary satellite. 
With this solution the equation of motion 
(geodesic equation) predicts that the satellite will fall down.
What is missing are the centrifugal forces, which arise
from our choice of Earth-fixed axes.


We now can do one of two things: 
1) Either we 
    {\it do not admit} 
coordinate systems which  
(in the asymptotic region for the solar system) 
are rotating relative to quasars. Then the usual asymptotic condition
$g_{\mu \nu} \rightarrow \eta_{\mu \nu}$ is adaequate. 
But excluding a class of coordinate systems 
is against the spirit of general relativity.
2) Or we 
    {\it do admit} 
coordinate systems which are rotating
asymptotically relative to quasars, e.g. we admit
coordinate systems with Earth-fixed axes.
But then we 
    {\it need encoding boundary conditions} 
in the asymptotic part of the solar system.
The boundary conditions 
    {\it encode fictitious forces}
(e.g. the Coriolis force, which is equivalent to a homogeneous
gravitomagnetic field) in the asymptotic part of the solar system.
In other words the boundary conditions 
encode the effects of sources outside the system considered 
(outside the solar system), 
they encode the influence of the cosmological sources.


The necessity of boundary conditions at spatial infinity
is familiar from electromagnetism. 
Boundary conditions are needed for 
fields, whose sources lie outside the system under discussion. 
In ordinary magnetism the sources external to the system
could be coils with electric currents 
generating applied external magnetic fields
for the experiment under consideration.


What is the general form of the boundary conditions 
in the region asymptotic to the solar system? 
The forms of the asymptotic force fields are 
familiar from the
fictitious forces in classical mechanics arising in coordinate
systems which are rotating relative to inertial systems.
In the vector sector of linear perturbations (gravitomagnetism)
two fictious forces arise, since  
the centrifugal force
$m[ \vec{\Omega} \times [\vec{r} \times \vec{\Omega}]]$ 
is absent at the level of linear perturbation theory:


1) The Coriolis force
$2m[\vec{v} \times \vec{\Omega}],$ 
which corresponds to a homogeneous gravitomagnetic field, 
\begin{equation}
\vec{B}_{{\rm g}} (\vec{x}, t) \rightarrow 
(\vec{B}_{{\rm g}})^{\, {\rm asy}}_{\, {\rm hom}} 
= - 2 \, \vec{\Omega}_{\, {\rm gyro}}^{\, {\rm asy}},
\end{equation}
where $\vec{\Omega}_{{\rm gyro}}^{\, {\rm asy}} $ is the 
precession vector of the spin axes of asymptotic gyroscopes 
(i.e. axes of inertial frames)
relative to the axes of our FIDO's
(e.g. Earth-fixed axes).


2) The fictitious force due to a non-constant angular velocity,
           $m[\vec{r} \times d \vec{\Omega}/dt],$ 
which corresponds to a gravitoelectric induction field,
\begin{equation} 
\vec{E}_{{\rm g}} (\vec{x}, t) \rightarrow
(\vec{E}_{{\rm g}})^{\, {\rm asy}} 
= [\vec{r} \times \frac{d}{dt} \, 
\vec{\Omega}_{\, {\rm gyro}}^{\, {\rm asy}} ].
\end{equation}
The asymptotic $\vec{B}_{\rm g}$ and 
               $\vec{E}_{\rm g}$ fields
are solutions of the source-free Einstein equations, 
i.e. they do not have any sources within our system
(solar system). 
Mach's principle implies that the sources for these fields
are given by cosmological matter 
(which is at infinity from the point of view of 
the solar system), as we have shown in 
    Eqs.~(\ref{B.general}, \ref{Faraday}).


One must distinguish two types of boundary conditions:
1) Boundary conditions which allow to 
{\it include and encode} the contribution
of sources outside the system considered. 
2) Boundary conditions which {\it exclude} unphysical solutions, 
e.g. the exponentially growing solution of
   Eq.~(\ref{Ampere.FRW.A}).
Boundary conditions discussed 
in the context of Mach's principle
refer to the encoding type.


In this paper we have shown that in 
     {\it cosmological gravitomagnetism} there is 
     {\it no need for boundary conditions of the encoding type}, 
because of the exponential cutoff for the contributions 
by the measured energy-momentum tensor of matter.


Einstein and others had hoped for some time
that one could implement the ideas of Mach
by specifying some boundary conditions 
on the equations of General Relativity.
But in 1934 Einstein   
    \cite{Einstein.closed}
concluded that the 
problem of boundary conditions (for Mach's principle)  
could only be solved by going to a spatially closed universe:
\begin{quotation}
``In my opinion the general theory of relativity 
can only solve this problem [of inertia] satisfactorily 
if it regards the world as spatially self-enclosed.''
\end{quotation}
In contrast we have shown in this paper that General Relativity for 
linear cosmological perturbations of a 
     {\it spatially flat} FRW universe
satisfies Mach's principle 
     {\it without the need for boundary conditions of the encoding type}. 
In a companion paper 
    \cite{my.FRW.not.spatially.flat}
we shall show that this is also true 
for perturbations of FRW universes with $k= -1,$
i.e. for open, spatially hyperbolic universes.


\section{Mach's Principle in the Presence of Scalar and Tensor Perturbations}
\label{sect.scalar.tensor}


Linear scalar perturbations cause 
weak gravitational lensing.
The geodesics from the gyroscope to quasars (on $\Sigma_t)$ 
will suffer geodesic deviation. 
This forces us to adapt the method of fixing the spatial axes 
of our FIDO's,
i.e. our LONB's, as follows: 
We demand that the spatial axes
of the LONB's do not rotate relative to the 
{\it average} of the directions of geodesics (on $\Sigma_t)$
to distant quasars, 
where the weight function in the average is the
{\it toroidal vector field} 
$\vec{X}^-_{\ell =1, m} = -i \vec{L} Y_{\ell =1, m} \, $
     \cite{my.FRW.not.spatially.flat}, 
which is tangential to the 2-sphere, 
\begin{equation}
0 = \int dr \, W(r)   
\int  d\Omega \, 
(\vec{X}^-_{\ell=1,m})^* \cdot 
\vec{v}_{\, {\rm asy}}^{\, \rm tang}.
\label{average.for.LONB}
\end{equation}
The FIDO at the origin determines 
$\vec{v}_{\, {\rm asy}}^{\, \rm tang}$
from the measured angular velocities of tangents 
to geodesics
on $\Sigma_t$ 
from the origin to quasars in the asymptotic FRW universe.
An observational window function in the asymptotic FRW universe
is denoted by $W(r).$


The vector field of distortions by weak gravitational lensing
belongs to the scalar sector.
This vector field is orthogonal
(under integration over the 2-sphere)
to the toroidal vector spherical harmonics $\vec{X}^-_{\ell=1,m},$ 
because the two fields belong to the opposite parity sequence.
The scalar sector has 
 $P=(-1)^{\ell};$ 
the toroidal vector spherical harmonics have
 $P=(-1)^{\ell +1}$.
Therefore 
     {\it scalar perturbations cannot affect Mach's principle}, 
if the gyroscope axes are
measured relative to the 
     {\it average}
of the directions of geodesics to
asymptotic quasars with the weight function 
$\vec{X}^-_{\ell =1, m}.$


Gravitational waves at the linear level 
in the perturbed region 
     {\it between} 
asymptotic FRW space and the gyroscope also
cannot affect the average of geodesics in
    Eq.~(\ref{average.for.LONB}), 
again because the tensor sector is orthogonal to the vector sector.


Gravitational waves 
    {\it going through} 
the gyroscope
cannot produce a gyroscope precession 
(relative to quasars in the asymptotic unperturbed universe).
This is because the precession of the gyroscope only depends 
on the three independent rotation coefficients
$(\omega_{\hat{i} \hat{j}})_{\hat{0}}$ 
for a displacement of unit measured time 
along the worldline of the gyroscope.
These three rotation coefficients form an axial 3-vector,
which cannot be affected by a 3-tensor (gravitational wave sector) 
for reasons of symmetry under 3-rotations.


If angular velocities are 
   {\it measured relative to the gyroscope axes} 
(not relative to distant quasars), 
one has the value
$\vec{\Omega}_{\rm gyro} = 0$ on the left-hand side of
   Eq.~(\ref{B.general}),
and there is no need to
modify the formula due to scalar or tensor perturbations,
because
    Eq.~(\ref{B.general})
automatically performs a projection on 
$\{\ell =1,$ odd parity sequence$\}$.


Einstein in his paper ``Prinzipielles zur 
allgemeinen Relativit\"atstheorie'' of 1918 
    \cite{Einstein.Mach}
coined the term ``Mach's Principle''
and gave the following formulation for it:
\begin{quotation} 
``The $G$-field is entirely determined ...  
by the energy tensor of matter.''
\end{quotation}
The $G$-field refers to the metric tensor field $g_{\mu \nu} (x).$ 


For linear cosmological perturbations 
Einstein's formulation is 
too strong to be valid, 
and  
unnecessarily strong for Mach's original purpose 
of explaining (giving the physical cause for) 
the evolution of groscope axes (i.e. non-rotating frames): 
1)~{\it Too strong}, as has been stated many times, 
because $g_{\mu \nu}$ can contain 
gravitational waves, which 
can exist without matter being present.
2)~{\it Unnecessarily strong}, as we shall now explain:
Gravitational waves 
cannot affect gyroscope axes {\it directly},
because a tensor perturbation cannot affect the spin evolution,
$d \vec{S}/dt,$ which is an axial vector.
And gravitational waves cannot affect the
$(\vec{X}^-_{\ell =1, m})$-average 
of the geodesics from a gyroscope to the asymptotic
unperturbed quasars.


In our approximation of {\it linear} cosmological perturbations
a third question remains unaddressed: 
Since the gyroscope's precession is determined by the
$\vec{B}_{\rm g}$ field at the gyroscope's position,
is $\vec{B}_{\rm g}$ determined 
by an equation which has 
a gravitational pseudo-angular-momentum tensor 
(in addition to the angular momentum of matter)
as a source term? 
Such an equation could not be Einstein's equation
in the standard way of separating 
output curvature terms from the matter input.


\section{Outlook}


%
\begin{enumerate}
\item The analysis of this paper has been extended to FRW universes 
      with $k =+1$ (spatially closed, spherical) and 
      with $k =-1$ (spatially open, hyperbolic)
      with analogous results.
      This will be presented in a companion paper
      \cite{my.FRW.not.spatially.flat}. 
\item Causality and astronomical observations 
      make a formulation desirable,
      where $\vec{B}_{\rm g}$ at the space-time point $P$ 
      is given by an integral 
      over sources on the 
         {\it backward light-cone} 
      of $P$. 
\item    {\it Second order perturbation theory:}
      Does perfect dragging of gyroscope axes still hold?
      What is the contribution of 
         {\it gravitational waves}
      to dragging of gyroscope axes?---  
      Can one derive 
         {\it exact results} 
      about perfect dragging of gyroscope axes? 
\item Does the Bianchi IX model of Ozsv\`ath and Sch\"ucking 
      satisfy Mach's principle
      (which involves an integral over energy flows in the universe)?
\item Accelerated frames of reference include not only 
      rotating frames (extensively discussed by Mach), 
      but also frames with linear acceleration 
      relative to inertial frames. 
      Does cosmological matter (with an exponential cutoff)
      cause 
      {\it perfect frame dragging} with respect to
      {\it linear accelerations?}  
\end{enumerate}
%


\begin{acknowledgements}
We would like to thank
E. Bertschinger, A. Guth, 
J. Hartle, 
J. Peebles, P. Steinhardt,
K. Thorne, and
R. Wald 
for discussions and hospitality, 
J. Ehlers, J. Hwang, D. Schwarz, and N. Straumann for discussions.
We would like to thank 
T. Cereghetti for collaboration 
during the work on his diploma thesis 
in an early stage of this project. 
\end{acknowledgements}


\bibliography{MachPR1.bbl}  

\end{document}